\documentclass[floatfix,groupedaddress,superscriptaddress,aps,showpacs,amsmath,amssymb,floatfix, prl, longbibliography, twocolumn,a4]{revtex4-1}

\usepackage{graphicx,float}
\usepackage{subfigure}
\usepackage{dcolumn}
\usepackage{bm}
\usepackage{mathrsfs}
\usepackage{txfonts}
\usepackage{CJK}
\usepackage[amssymb]{SIunits}
\usepackage{epsfig}
\usepackage{epstopdf}
\usepackage{lipsum}
\usepackage{color}
\usepackage{placeins}

\newenvironment{SChinese}{%
\CJKfamily{gbsn}%
\CJKtilde
\CJKnospace}{}

\allowdisplaybreaks[2]

\begin{document}

\begin{CJK}{UTF8}{}
\begin{SChinese}

\title{On-chip chiral single-photon interface: Isolation and unidirectional emission}

 \author{Lei Tang}
 \affiliation{National Laboratory of Solid State Microstructures, College of Engineering and Applied Sciences, and School of Physics, Nanjing University, Nanjing 210093, China}

 \author{Jiangshan Tang}
 \affiliation{National Laboratory of Solid State Microstructures, College of Engineering and Applied Sciences, and School of Physics, Nanjing University, Nanjing 210093, China}

  \author{Weidong Zhang}
 \affiliation{State Key Laboratory for Mesoscopic Physics $\&$ Collaborative Innovation Center of Quantum Matter, Department of Physics, Peking University, Beijing 100871, China}

 \author{Guowei Lu}
 \affiliation{State Key Laboratory for Mesoscopic Physics $\&$ Collaborative Innovation Center of Quantum Matter, Department of Physics, Peking University, Beijing 100871, China}

  \author{Han Zhang}
 \affiliation{Collaborative Innovation Center of Advanced Microstructures, and School of Physics, Nanjing University, Nanjing 210093, China}

 \author{Yong Zhang}
 \email{zhangyong@nju.edu.cn}
 \affiliation{National Laboratory of Solid State Microstructures, College of Engineering and Applied Sciences, and School of Physics, Nanjing University, Nanjing 210093, China}

 \author{Keyu Xia}
 \email{keyu.xia@nju.edu.cn}
 \affiliation{National Laboratory of Solid State Microstructures, College of Engineering and Applied Sciences, and School of Physics, Nanjing University, Nanjing 210093, China}

 \author{Min Xiao}
 \affiliation{National Laboratory of Solid State Microstructures, College of Engineering and Applied Sciences, and School of Physics, Nanjing University, Nanjing 210093, China}
 \affiliation{Department of Physics, University of Arkansas, Fayetteville, Arkansas 72701, USA}

\date{\today}

\begin{abstract}
 Chiral quantum systems have received intensive attention in fundamental physics and applications in quantum information processing including optical isolation and photon unidirectional emission. Here, we design an on-chip emitter-resonator system with strong chiral light-matter interaction for a chiral single-photon interface. The system includes a microring resonator with a strong evanescent field and a near-unity optical chirality along both of the whole outside and inside walls, allowing a strong and chiral coupling of the Whispering-Gallery mode to a quantum emitter. By initializing a quantum dot in a specific spin ground state or shifting the transition energy with a polarization-selective optical Stark effect, we show a broadband optical isolation at the single-photon level over several \giga\hertz. Furthermore, a quantum emitter chirally coupling to the microring resonator can emit single photons unidirectionally. Our protocol  paves a way to realize multifunctional chiral single-photon interface in on-chip quantum information processing and quantum networks.
\end{abstract}


\maketitle

\end{SChinese}
\end{CJK}


 Optical chirality, non-reciprocity and unidirectional emission are of particular interest in the fundamental science \cite{PNAS.113.6845,  Science.358.636, Science.359.4003, Science.356.1260,  Science.359.1231, plasmon1, nphoton.9.796, PRL.112.043904, KeyuChiralKerr} and promise important applications in modern optical systems \cite{votexemitter} and quantum information processing \cite{lodahl2017chiral, PRL.113.237203, PRL.117.240501, PRL.115.163603, PRA.97.062318}. The recent progress in these fields has led to an emerging field called ``chiral quantum optics" \cite{lodahl2017chiral,nanofiber2, bottle2circulator, WG-deterministicemitter1, WG-deterministicemitter2, Science.359.666, PRL.115.153901, PRL.120.043901, pcwaveguide2, PRB.95.121401, Xia.2014, KeyuNaturePhoton, KeyuChiralKerr}.

 A strong chiral light-matter interaction is the basis of chiral quantum optics and achieved by coupling a quantum emitter (QE) with photon-spin dependent transitions to an electric (e-) field, transversely confined in a subwavelength space and consequently possessing the  ``spin-moment locking'' (SML) at particular positions \cite{lodahl2017chiral, Science.348.1448, bliokh2015transverse,  bottle2circulator, Luis.nanofiber, nphoton.9.789, PRL.102.033902, PRB.95.121401, PRL.107.173902,PRL.113.093603, Xia.2014}. Realizing chiral light-matter interaction require either the magnetic-field-induced Zeeman shift \cite{WG-deterministicemitter1} or an asymmetric dipole moment \cite{Science.348.1448, PRX.5.021025, Xia.2014}.
 This letter will focus on proposing a novel chiral interface for single photons by initializing a QE in a special spin state or using the optical Stark control.

 Although optical non-reciprocity has been well studied in various systems and using different scenarios \cite{nphoton.3.91, nphysics.10.394, nphoton.8.524, Nat.Comm.7.13657, PRL.120.203904, PRL.118.033901, PRX.5.021025, PRL.110.093901, PRL.110.223602, nphoton.11.774, PNASChen, PRA.97.013802, Science.333.729, PRL.121.153601, Nat.558.569, JiangXiaoShunNonRecpPT}, optical isolation at the single-photon level has only been reported in quantum optical systems with chiral light-matter interaction, based on the photonic SML \cite{bottle2circulator, nanofiber2, WG-deterministicemitter1, Xia.2014, Science.345.903} or the photonic Aharonov-Bohm effect \cite{OL.40.5140}. The chiral-waveguide-based or chiral-cavity-based single-photon isolation normally has a narrow bandwidth, typically up to tens of $\mega\hertz$ \cite{bottle2circulator, nanofiber2, Xia.2014}, limited to the edge of the band or the weak evanescent e-field due to a large transverse dimension \cite{WG-deterministicemitter1, PRL.115.153901, Xia.2014, PRL.113.093603, bottle2circulator, nanofiber2, Opt.Express.21.23942}. Additionally, the QE needs to be positioned precisely in a nanosize region.  Moreover, unidirectional emission of single photons is highly desired but has only been demonstrated in a chiral waveguide-emitter system \cite{Science.359.666, WG-deterministicemitter1, pcwaveguide2, PRB.95.121401}.

 In this letter, we present a CMOS-compatible chiral photonic interface for single-photon isolation and unidirectional emission. In our design, the silicon microring resonator with a subwavelength transverse dimension has an exceptionally strong evanescent e-field and a unity optical chirality (OC) surrounding the whole outside and inside walls. Therefore, even the resonator with a moderate quality factor $10^4$ can strongly couples to a negatively charged quantum dot (QD) in a chiral way.  In this, we can realize broadband single-photon isolation, and achieve unidirectional and polarization-deterministic single-photon emission.


  \begin{figure}
  \centering
  \includegraphics[width=0.8\linewidth]{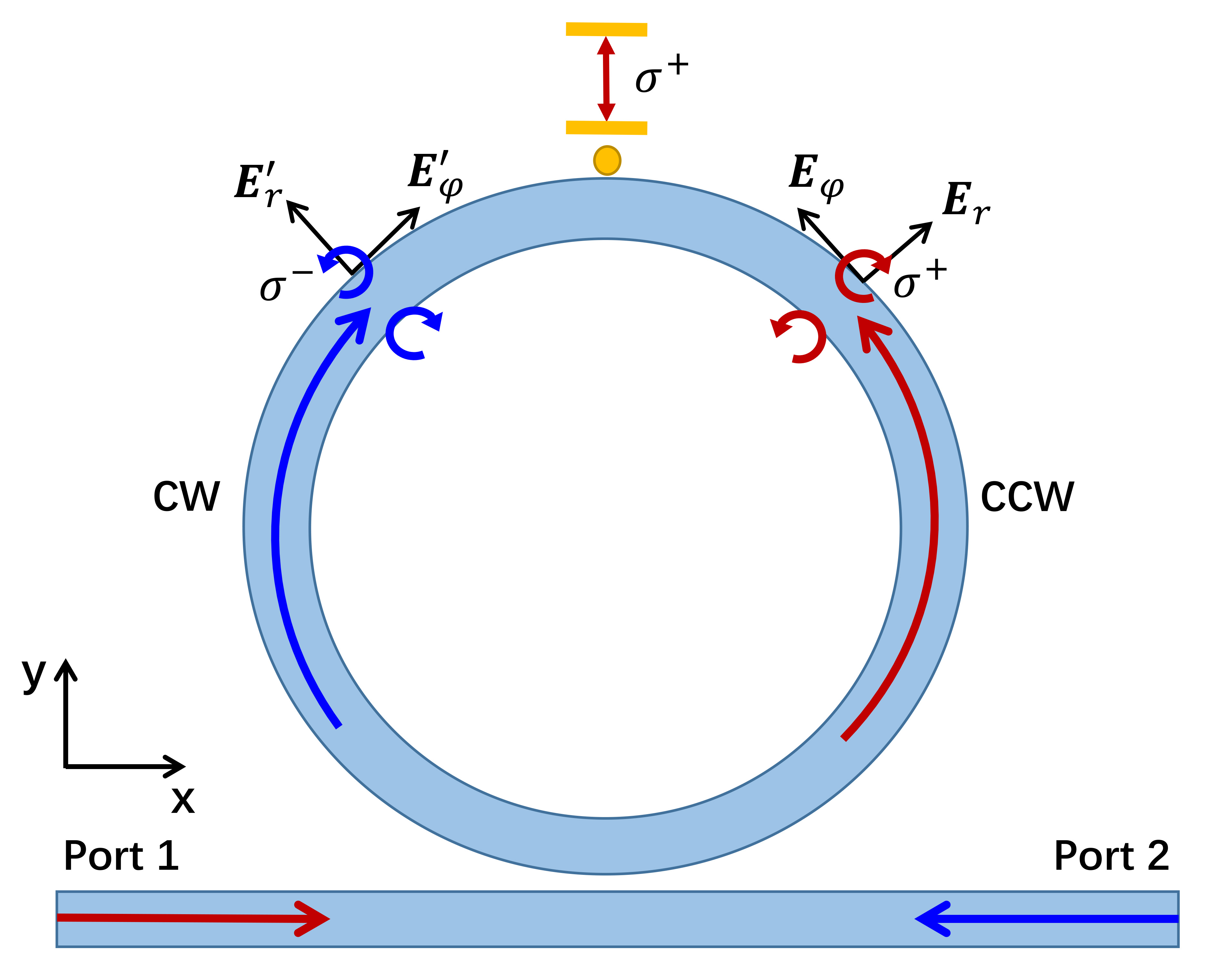} \\
  \caption{Schematic of the chiral quantum optical system. A microring resonator couples to a nearby waveguide and a single negatively charged QD. The light incident to the port $1$ ($2$) drives the CCW (CW) mode.  The polarization of the evanescent field of the CCW mode is  $\sigma^+$- ($\sigma^-$-) polarized near the whole outside (inside) wall, while that for the CW mode is  $\sigma^-$- ($\sigma^+$-) polarized. The QD is treated as a  two-level system with $\sigma^+$-polarized transition.}
  \label{fig:FIG1}
  \end{figure}
  %
  Our QD-resonator system, depicted in Fig.~\ref{fig:FIG1}, consists of a silicon waveguide, a silicon microring resonator with the refractive index $n_1=3.48$, and a single negatively charged QD. The resonator and the waveguide are $0.44$~\micro\meter~wide and $0.22$~\micro\meter~thick. The resonator has a $4.22$~\micro\meter~radius. Its Whispering-Gallery modes (WGMs) decays into the waveguide at a rate $\kappa_\text{ex}$. Our numerical simulation with the finite-difference time-domain (FDTD) method yields an intrinsic quality factor of $Q_\text{in}\approx 3.9\times 10^4$ at the wavelength $\lambda_\text{c} \sim 1.556$~\micro\meter, and a mode volume $V_m\sim1.55~\micro\meter^3$.  The corresponding resonance frequency and the intrinsic decay rate are $\omega_\text{c}/ 2\pi \approx  192.67~\tera\hertz$ and $\kappa_\text{i}/ 2\pi\approx  4.94~\giga\hertz$, respectively, yielding a total decay rate of $\kappa = \kappa_\text{ex} + \kappa_\text{i} \approx 2\pi \times 9.88~\giga\hertz$. A higher Q factor, e.g. $Q\sim 10^5$ at  $ 1.55~\micro\meter$, has been experimentally demonstrated in a SOI mesoscopic resonator \cite{Qring1}, even for a smaller mode volume \cite{Qring2}. The relative low Q factor of our resonator is due to the large spatial grid in simulation, limited by available computation resource. The waveguide-resonator gap ($n_2=1$) is set to $0.19$~\micro\meter~ that the critical coupling condition is almost obtained, confirmed by a vanishing transmission, $T \sim 0$, of an empty resonator \cite{SupplMat1}.

Now we design the microring resonator that  the clockwise (CW) and counterclockwise (CCW) WGMs possess the SML. We numerically investigate the electric field distribution of these two modes. The input light from port $1$ or $2$ is almost exclusively transversally polarized, i.e. TE mode. Whereas the light circulating in the resonator is tightly confined in the transverse direction as a TM mode \cite{elliptical-evanescentwave,bottle1,votexemitter}. Thus, the evanescent e-field near the side surfaces of the resonator has a local longitudinal-polarizated component (${{\bf{E}}_\varphi }$) and a transverse component (${\bf{E}}_r$). These two components are $\pm {\pi  \mathord{\left/{\vphantom {\pi 2}}\right.\kern-\nulldelimiterspace} 2}$ out of phase with each other \cite{elliptical-evanescentwave}, with the $\pm$ sign depending on the propagating direction of the light (see Fig.~\ref{fig:FIG1}). The evanescent field of the WGM is inherently elliptically polarized with its polarization locked to the propagating direction.  The complex-valued amplitude of the evanescent field is given by $ {{\bf E}_\text{eva}} = {{\bf E}_r} \pm i {\bf{E}_\varphi} $.
  The ratio $|{\bf E}_\varphi | / |{\bf E}_r |$ can be estimated as $|{\bf E}_\varphi | / |{\bf E}_r | \approx \sqrt{1-(n_2/n_1)^2}$ \cite{bottle1}. In our design with $n_1=3.48$ and $n_2=1$, the ratio is about $0.96$. Thus, the evanescent fields are near perfectly circularly polarized, i.e. $\sigma^\pm$-polarized.

\begin{figure}
  \centering
  \includegraphics[width=1\linewidth]{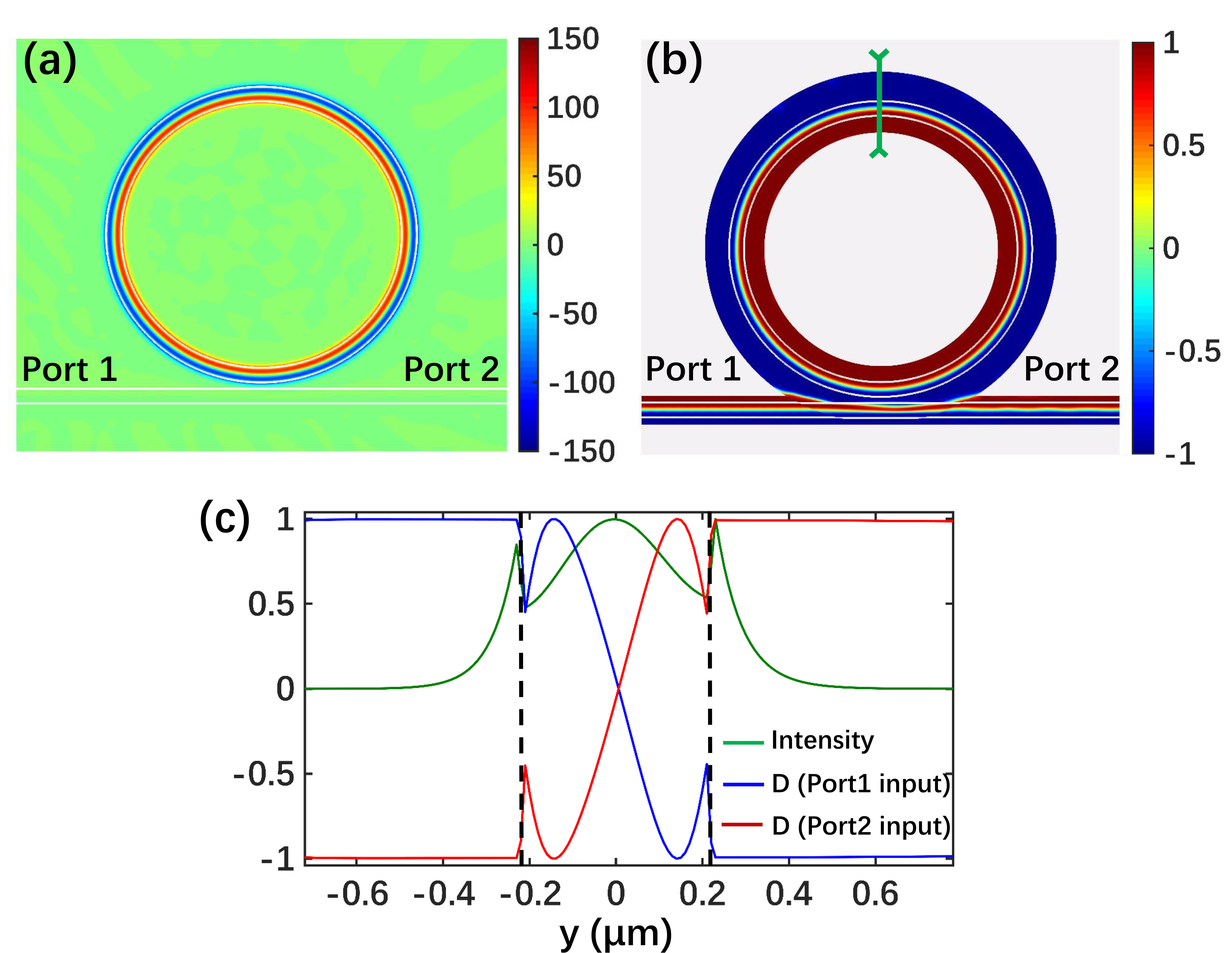} \\
  \caption{Optical chirality and field distribution. Intensity difference $C$ (a) and optical chirality $D$ (b) for light with $\lambda=1.55~\micro\meter$ incident to the port $1$. White lines show the waveguide boundaries. See the supplemental material for the case of the light incident to the port $2$. (c) Transverse distribution of the intensity difference $C$ and the OC $D$ at the position indicated by the green line in (b). Blue (red) curve describes $D$ for light incident to the port $1$ ($2$), and green curve describes the total electric field intensity. Black dashed lines show the resonator boundaries of the inside and outside walls at $y= - 0.22~\micro\meter$ and $y=0.22~\micro\meter$, respectively.}
  \label{fig:FIG2}
  \end{figure}

  Next, we numerically evaluate the OC of our resonator by FDTD simulation. We first calculate the intensity difference between the left-circularly ($\sigma^-$) and right-circularly ($\sigma^+$) polarized components, $C=(|{\bf{E}}({\bf{r}}) \cdot {{\bf{e}}_{{\sigma ^ - }}}{|^2} - |{\bf{E}}({\bf{r}}) \cdot {{\bf{e}}_{{\sigma ^ + }}}{|^2})$, at the position $\bf{r}$ with ${{\bf{e}}_{{\sigma ^ - }}}=({{\bf{e}}_x} - i{{\bf{e}}_y})/\sqrt 2$ and ${{\bf{e}}_{{\sigma ^ + }}}=({{\bf{e}}_x} + i{{\bf{e}}_y})/\sqrt 2$, where ${{\bf{e}}_x}$ and ${{\bf{e}}_y}$ are unit vectors along the $x$ and $y$ directions, respectively.
  For a TE mode input from the port 1, the intensity difference distribution $C$ is shown in Fig.~\ref{fig:FIG2}(a).
  The evanescent e-field along the outside (inside) wall is almost  $\sigma^+$- ($\sigma^-$-) polarized, while it is linearly-polarized in the middle of the resonator. The OC, defined as $D = (|{\bf{E}}({\bf{r}}) \cdot {{\bf{e}}_{{\sigma ^ - }}}{|^2} - |{\bf{E}}({\bf{r}}) \cdot {{\bf{e}}_{{\sigma ^ + }}}{|^2})/(|{\bf{E}}({\bf{r}}) \cdot {{\bf{e}}_{{\sigma ^ - }}}{|^2} + |{\bf{E}}({\bf{r}}) \cdot {{\bf{e}}_{{\sigma ^ + }}}{|^2})$ \cite{PRL.121.043901, PRL.104.163901,PRA.83.021803, OCDefinition},  figures in what degree the field is locked to the light momentum. It is an important value showing the chirality of the field.
  The value $D = 1 (-1)$ implies the field is entirely ${\sigma ^ - }$- (${\sigma ^ + }$-) polarized, while $D = 0$ corresponds to a linear polarization. Clearly, our designed resonator has nearly unity OC along both the outside and inside walls, see Fig.~\ref{fig:FIG2}(b). For example, when the light enters the waveguide from the port $1$ and excites the CCW mode, the outer (inner) evanescent e-field of the WGM is $\sigma^+$- ($\sigma^-$-) polarized, indicated by $D \approx -1 (1)$. More details of the fields and the OC are shown in Fig.~\ref{fig:FIG2}(c) for the cross area marked by the green line in Fig.~\ref{fig:FIG2}(b). In stark contrast to the linearly-polarized e-field at the middle of the resonator, the evanescent e-field is a nearly-perfect circular polarization for both cases of light incident to the port $1$ and port $2$. When the light is reversed, the polarization also interchanges. We obtain $|D| > 0.99$ from the surface of the outside wall to a position $280~\nano\meter$ away in the radial direction. This large chiral area greatly relaxes the requirement for precisely positioning a QD. Importantly, the intensities of the evanescent fields near the walls are almost equal to that in the middle of the resonator. This feature of our design, in comparison with the conventional bottle-shaped resonator, allows a stronger coupling between a nearby QD and the resonator.

 Now we go to describe the chiral and strong interaction between a QD and the resonator.  As shown in Fig.~\ref{fig:FIG1}, a negatively charged QD is positioned nearby the outside wall of the resonator. It has two energy-degenerate transitions at $\lambda_q \sim 1550~\nano\meter$, driven by a circularly-polarized e-field. Note that the light at this wavelength is transparent in silicon. It can be an InAs self-assembled  QD grown on the silicon-dioxide/silicon substrates \cite{QD1,QD2,QD3}, with two electronic spin ground states, $|1/2\rangle = | \uparrow\rangle$ and$|-1/2\rangle = | \downarrow\rangle$, and two optically excited states, $|3/2\rangle = |\uparrow \downarrow \Uparrow\rangle$ and $|-3/2\rangle = |\uparrow \downarrow \Downarrow\rangle$. The notation $|\Uparrow\rangle$ ($|\Downarrow\rangle$) denotes the spin-up (spin-down) hole state, and $|\uparrow\rangle$ ($|\downarrow\rangle$) is for the spin-up (spin-down) electronic state. When the QD is prepared in the $|1/2\rangle$ ($|-1/2\rangle$) state, it can only be driven by the $\sigma^+$- ($\sigma^-$-) polarized field to the state $|3/2\rangle$ ($|3/2\rangle$) in the absence of external magnetic field.  Initialization of the QD in either ground state has been experimentally demonstrated with a near-unity probability \cite{QDpreparation1, QDpreparation2, QDpreparation3}.  The polarization-selective transition, $|1/2\rangle \leftrightarrow |3/2\rangle$ or $|-1/2\rangle \leftrightarrow |-3/2\rangle$, can also be tuned to have different energies via the optical Stark effect (OSE) \cite{Nat.Commun.6.7665, Nat.Phys.14.1092, Nat.Phys.13.26, Sci.Adv.2.e1600477, Stark2, Stark3}. For simplicity, we assume that the QD is completely populated in the spin up ground state, or only allows the $\sigma^+-$polarized transition, enabled by the OSE. Thus, the QD can be treated as a two-level system with a $\sigma^+-$driven transition, see Fig.~\ref{fig:FIG1}. It only couples to the CCW WGM of the resonator. Note that the OSE-based method allows an all-optical operation. In fabrication, the QD can be engineered to have various resonance wavelengths,  dipole moments and decoherence rates. Here, we assume $\lambda_\text{q}\approx 1.556~\micro\meter$ that $\omega_\text{q} = \omega_\text{c}$ and a dipole moment $d=20~\text{Debye}$, yielding a spontaneous emission rate $\gamma_\text{q} = {d^2}\omega _\text{q}^2/3\pi {\varepsilon _0}\hbar {c^3} =  2\pi \times 5.29~\mega\hertz$ \cite{Luis.nanofiber}. Such parameters for the QD are experimentally available \cite{PRL.88.087401, APL.82.4552, Nat.Nanotechnol.13.398}. The mode volume is calculated to be $V_\text{m} = 1.55~\micro\meter^3$, yielding the QD-resonator coupling strength $g= d \sqrt{\omega_\text{c} / 2\epsilon_0 \hbar V_\text{m}} = 2\pi \times 6.86 ~\giga\hertz$. Thus, we reach the strong coupling regime, $g > \kappa, \gamma_\text{q}$, when $\kappa_\text{ex} = \kappa_\text{i}$.

 In our design, the QD strongly couples to the CCW WGM but decouples from the counter-propagating CW WGM. Therefore, our quantum QD-resonator system is chiral and subsequently allows one to realize the single-photon isolation. 
 Below we first investigate the single-photon isolation of our system with the single-photon scattering method, developed by Shen and Fan\cite{Fan, Opt.Lett.30.2001}. Then, we show the dynamic non-reciprocity with single photons input into the two ports simultaneously.
 We find the steady-state forward and backward transmission amplitudes, corresponding to the input to the port $1$ and port $2$ \cite{SupplMat2, Fan1, Fan, Opt.Lett.30.2001, Xia.2014}, respectively,
  \begin{subequations} \label{eq:T}
   \begin{widetext}
    \begin{align}
      {t_ + }  = & \frac{{\tilde{\Delta}_\text{c}\left[ {\tilde{\Delta}_\text{c} \tilde{\Delta}_\text{q} - \left( {{{\left| {{g_\text{a}}} \right|}^2} + {{\left| {{g_\text{b}}} \right|}^2}} \right)} \right] + \tilde{\Delta}_\text{q}{\kappa_\text{ex} ^2} - g_\text{a}^*{g_\text{b}}h - {g_\text{a}}g_\text{b}^*{h^*} - \tilde{\Delta}_\text{q}{{\left| h \right|}^2} +i \left( {{{\left| {{g_\text{b}}} \right|}^2} - {{\left| {{g_\text{a}}} \right|}^2}} \right)\kappa_\text{ex} }}{{\left( {\tilde{\Delta}_\text{c} + i\kappa_\text{ex} } \right)\left[ {\tilde{\Delta}_\text{q}\left( {\tilde{\Delta}_\text{c} + i\kappa_\text{ex} } \right) - \left( {{{\left| {{g_\text{a}}} \right|}^2} + {{\left| {{g_\text{b}}} \right|}^2}} \right)} \right] - g_\text{a}^*{g_\text{b}}h - {g_\text{a}}g_\text{b}^*{h^*} - \tilde{\Delta}_\text{q}{{\left| h \right|}^2}}} \;, \\
      {t_ - }  = & \frac{{\tilde{\Delta}_\text{c}\left[ {\tilde{\Delta}_\text{c} \tilde{\Delta}_\text{q} - \left( {{{\left| {{g_\text{b}}} \right|}^2} + {{\left| {{g_\text{a}}} \right|}^2}} \right)} \right] + \tilde{\Delta}_\text{q}{\kappa_\text{ex} ^2} - g_\text{b}^*{g_\text{a}}h - {g_\text{b}}g_\text{a}^*{h^*} - \tilde{\Delta}_\text{q}{{\left| h \right|}^2} +i \left( {{{\left| {{g_\text{a}}} \right|}^2} - {{\left| {{g_\text{b}}} \right|}^2}} \right)\kappa_\text{ex} }}{{\left( {\tilde{\Delta}_\text{c} + i\kappa_\text{ex} } \right)\left[ {\tilde{\Delta}_\text{q}\left( {\tilde{\Delta}_\text{c} + i\kappa_\text{ex} } \right) - \left( {{{\left| {{g_\text{b}}} \right|}^2} + {{\left| {{g_\text{a}}} \right|}^2}} \right)} \right] - g_\text{b}^*{g_\text{a}}h - {g_\text{b}}g_\text{a}^*{h^*} - \tilde{\Delta}_\text{q}{{\left| h \right|}^2}}} \;,
    \end{align}
   \end{widetext}
  \end{subequations}
  where $\tilde{\Delta}_\text{c} = \omega  - \omega_\text{c}  + i{\kappa_\text{i}}$ and $\tilde{\Delta}_\text{q} = \omega  - \omega_\text{q}  + i{\gamma _\text{q}}$; $\kappa_\text{ex} = V^2/2v_\text{g}$ is the external decay rate of the resonator due to the coupling $V$ to the waveguide; and ${v_\text{g}}$ is the group velocity of the photon in the waveguide. $g_\text{a}$ ($g_\text{b}$) is the coupling strength between the CCW (CW) WGM and the QD, $h$ models the intermode backscattering between the CCW and CW WGMs, typically due to the surface roughness. We define the detuning $\Delta_\text{c}  = \omega - \omega_\text{c}$ and always assume ${\omega _\text{c}} = \omega_\text{q}$. The forward and backward transmissions are $T_+ = {\left| t_+ \right|^2}$ and $T_- = {\left| t_- \right|^2}$, respectively. We have $|g_\text{a}| = g\sqrt{(1-D)/2}$ and $|g_\text{b}| = g\sqrt{(1+ D)/2}$.

\begin{figure}[h]
  \centering
  \includegraphics[width=1\linewidth]{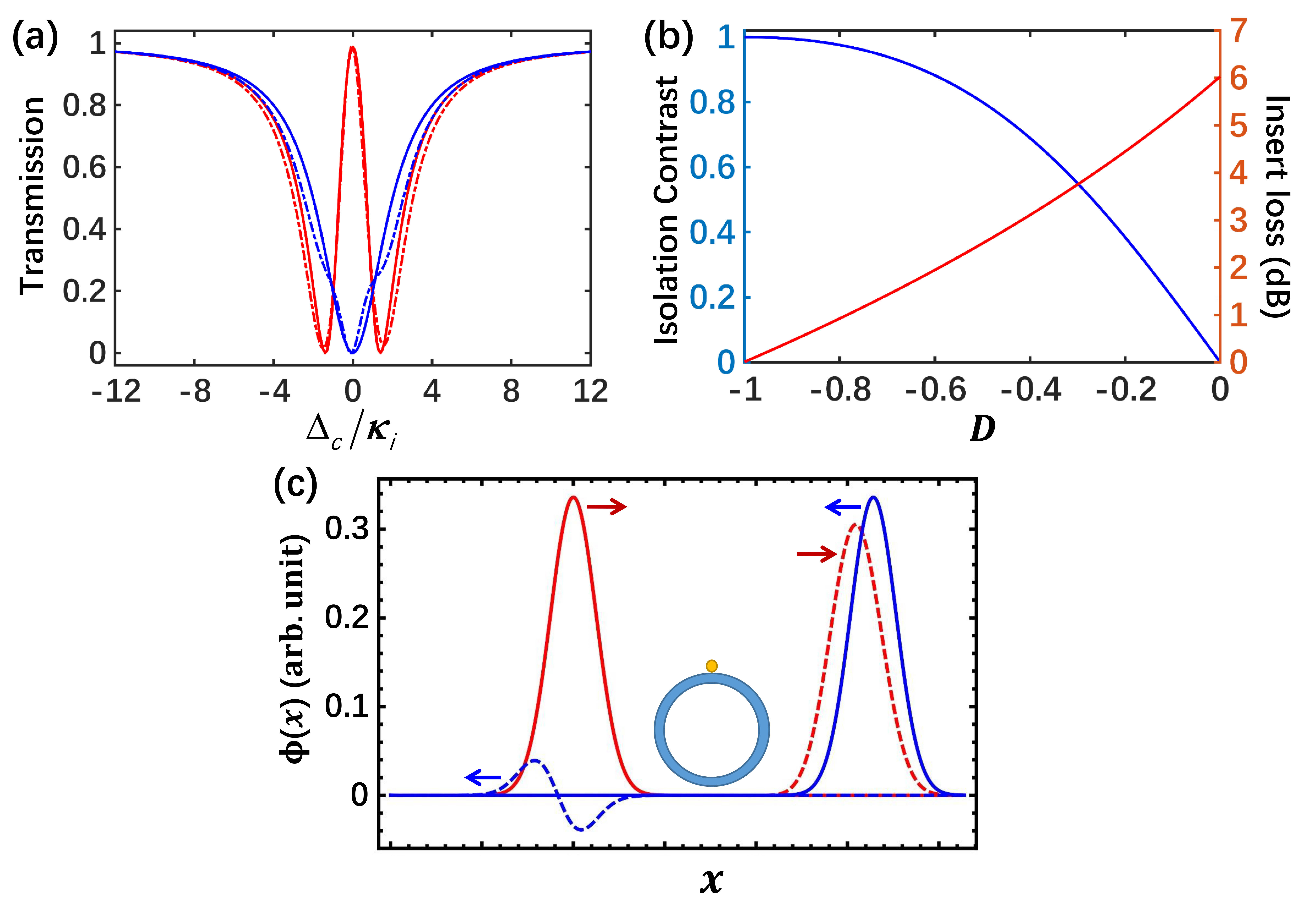} \\
  \caption{Single-photon transmission. (a) Steady-state transmissions for $|D|=0.99$. Red (blue) curves are for the forward (backward) transmissions ${T_ +}$ ($T_-$), without the backscattering, i.e. $h=0$, (solid curves) and with a backscattering of $|h|=\kappa_\text{i}$ (dashed curves). (b) Isolation contrast (blue curve) and insertion loss (red curve) as a function of the OC $D$  for $h=0$.  $\Delta_\text{c} = \Delta_\text{q}=0, g \approx 1.39 \kappa_\text{i}, \gamma_\text{q} =10^{-3} \kappa_\text{i}$, and $\kappa_\text{ex}=\kappa_\text{i}$. (c) Propagation of single-photon pulses incident to the port $1$ and port $2$ simultaneously. Red (blue) curves show the forward (backward) propagation of single-photon pulses input to the port $1$ ($2$). Solid curves are the input single-photon wavefunction, and dashed curves for the transmitted wavefunction. Other parameters are as in (a) and (b) but $D = -1$ for simplicity. }
  \label{fig:FIG3}
  \end{figure}

  The steady-state forward and backward transmissions for different detunings and OCs are shown in Fig.~\ref{fig:FIG3}(a) and (b). For our special design,  we have $D = - 0.99$ and $|h| \ll \kappa_\text{i}$, confirmed by the singlet peak at $1.556~\micro\meter$ of the transmission of the bare resonator without  the QD \cite{SupplMat3}. The performance of the single-photon isolation for $D = - 0.99$ is shown in Figs.~\ref{fig:FIG3}(a). In the absence of the backscattering, i.e. $h=0$, we obtain $T_+ \approx 0.99$ and $T_- \approx 0$ at $\Delta_\text{c} = 0$, corresponding to the insertion loss of $\mathscr{L} = -10 log(T_+) \approx 0.04~\text{dB}$ and the isolation contrast $\eta = (T_+ - T_-)/(T_+ + T_-) \approx 1$ \cite{PRL.107.173902, Xia.2014}. Obviously, the single-photon isolation is achieved with almost zero insert loss and near-unity isolation contrast.  Even for a relatively large backscattering $|h|=\kappa_\text{i}$, both the forward and backward transmissions only change very slightly, meaning a very small reduction in the performance. The nonreciprocal bandwidth is about $0.7 \kappa \approx 2\pi \times 7~\giga\hertz$, limited by the available QD-resonator coupling strength. To our best knowledge, this spectral window is about two-to-three orders broader than the previous achievements \cite{nanofiber2, bottle2circulator, WG-deterministicemitter1, KeyuNaturePhoton,KeyuChiralKerr}. As seen from Fig.~\ref{fig:FIG3}(b), the isolation contrast is quite robust, decreasing slowly from $1$ to $0.8$ as the OC decreases from $-1$ to $-0.5$. While the insertion loss increases almost linearly during this region.

 Many previous schemes for optical isolation suffer the dynamic reciprocity problem when oppositely propagating lights enter the system at the same time \cite{Nat.Photonics.9.388}. Our scheme can circumvent this challenging problem. To prove this point, we perform numerical simulations for the propagation of single-photon wave packets incident to the port $1$ and port $2$ simultaneously \cite{SupplMat4, Fan1, Opt.Lett.30.2001}, as shown in Fig.~\ref{fig:FIG3}(c). We set the velocity of light in the waveguide $v_g=1$, and apply the critical coupling condition. We apply Gaussian single-photon pulses with a bandwidth of $0.2 \kappa$. At resonance, the right-moving single-photon can pass through the system with a transmission $T_+ = 0.91$. In contrast, the backward transmission probability of a left-moving single photon is only $0.02$.

  \begin{figure}[h]
  \centering
  \includegraphics[width=1\linewidth]{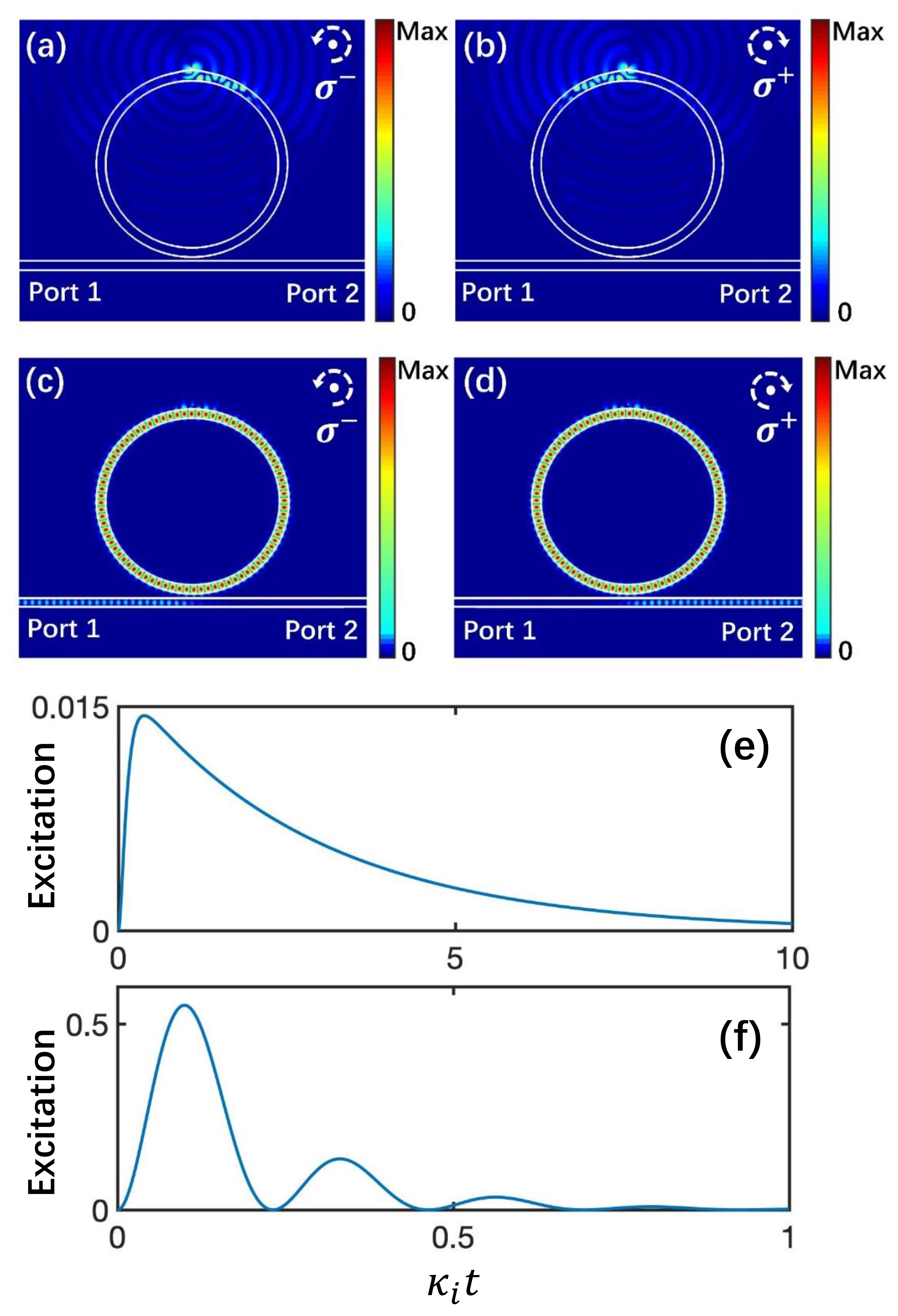} \\
  \caption{(a-d) Magnetic field distributions for a circularly-polarized dipole coupling to the microring resonator. Left (right) panels show the magnetic field for a Gaussian-shaped $\sigma^-$- ($\sigma^+$-) polarized dipole with a duration of $\tau_\text{p}=110~\pico\second$. (a) and (b) for the instantaneous magnetic field $\left| {{\mathop{\rm Re}\nolimits} \left( H \right)} \right|$ (for a higher visibility of the field in the waveguide) at $t=30~\femto\second$. (c) and (d) for $\left| {{\mathop{\rm Re}\nolimits} \left( H \right)} \right|$ at $t=120~\pico\second$. (e) Excitations of the waveguide mode for $\kappa_\text{i} = 2\pi \times 4.94$~\giga\hertz~ ($Q_\text{in} \sim 3.9 \times 10^4$), and $\kappa_\text{ex} = 10 \kappa_i$. (f) as in (e) but $\kappa_\text{i} = 2\pi \times 0.494$~\giga\hertz, and $\kappa_\text{ex} = 5 \kappa_i$. Other parameters in (e) and (f) are $\omega_\text{c}=\omega_\text{q}, \gamma_\text{q} = 10^{-3}\kappa_\text{i}$ and $g = 2\pi \times 6.86$~\giga\hertz~fixed.}
  \label{fig:FIG4}
  \end{figure}

 When the QD is initially prepared in its excited state, it will emit a single photon into either the CW resonator mode or the CCW one in the strong coupling regime. The exiting path of the photon is determined by the populated excited state of the QD. Therefore, by initializing the QD in a spin-selective excited state, we can realize the unidirectional emission of single photons with a deterministic polarization (an eigenmode of the waveguide). We are interested in the emission direction of photons. Thus, we replace the QD with a circularly-polarized Gaussian-pulse electric dipole, $E_\text{d}(t) = \pi^{-1/4} \tau_\text{p}^{-1/2} \exp( - {(t - {\tau _\text{d}})^2}/2\tau _\text{p}^2) \sin(\omega_\text{c} (t - {\tau _\text{d}}))$, in the FDTD simulation, where $\tau_\text{p}$ is the duration of the dipole-emitted photon pulse, and $\tau_\text{d}$ the delay.  When the QD is prepared in the state $|- 3/2\rangle$ corresponding to a $\sigma^-$-polarized dipole, it exclusively excites the CW mode, see Figs.~\ref{fig:FIG4} (a). The emitted single photon exits the system through the port $1$, as shown in Figs.~\ref{fig:FIG4} (c). When the state $|3/2\rangle$ is initially populated (given a $\sigma^+$-polarized dipole), the CCW mode is excited, and the single photon comes out from the port $2$ instead, see  Figs.~\ref{fig:FIG4} (b) and (d). The dipole is on resonance with the WGM at $\lambda_\text{c} = 1.556~\micro\meter$ and $\tau_\text{p} > 2\pi/\kappa$. We numerically solve the quantum Langevin equations for calculating the single-photon excitation collected by the waveguide \cite{SupplMat5, toolbox}. For a low-Q cavity with $Q_\text{in} \sim 3.9 \times 10^4$ and $\kappa_\text{ex} =10 \kappa_\text{i}$, the emitted long-pulsed single photon is captured by the resonator and then is collected with an excitation of $\sim 0.91$ by the waveguide, see Fig.~\ref{fig:FIG4}(e).  Such unidirectional single-photon emission with a deterministic polarization (TE eigenmode of the waveguide)  is important for scalable quantum computation but challenging \cite{ChaoyangLu}. If the cavity intrinsic Q-factor can reach $Q_\text{in} \sim 3.9 \times 10^5$ (already available experimentally \cite{Qring1}), a time-bin single photon, useful in quantum information technologies \cite{PRL.82.2594, Nat.Commun.5251}, is obtained with a total excitation of $0.83$ [Fig.~\ref{fig:FIG4}(f)]. 

Photon blockade can be achieved in a strongly-coupled QE-cavity system \cite{ Science.345.903, Science.319.1062}.  A nonreciprocal version has only been proposed recently with a fast spinning resonator \cite{PRL.121.153601}. Because the QD strongly couples to the CCW GWM but decouples from the CW one, our solid-state device can also perform nonreciprocal photon blockade without moving parts. 

 In conclusion, we have proposed a chiral single-photon interface with a QD-resonator system. The evanescent e-field of the resonator is strong and perfectly circularly-polarized along the whole side surfaces. Thus, the resulting strong light-matter interaction with a near-unity OC can be achieved without the requirement of precisely positioning the QD as the previous works. We further show a \giga\hertz-bandwidth single-photon isolator and controllable unidirectional emission of single photons. Our protocol can be extended to a chiral quantum system consisting of a subwavelength resonator interacting with 2D material or perovskites, prepared and operating at room temperature \cite{Nat.Phys.13.26, Sci.Adv.2.e1600477, Nat.Phys.14.1092, Science.359.443, valleytronics, Nat.Phys.13.894}.  It provides an on-chip platform for a multifunctional single-photon interface.

 The authors thank H.-D. Wu, Y.-G. Liu, Dr. T. Li for helpful discussions. H.Z., Y.Z., K.X. and M.X. thank the support of the National Key R\&D Program of China (Grant No. 2017YFA0303703). This work is also supported by the National Natural Science Foundation of China (Grant Nos. 11874212, 11574145).


%

\end{document}


\preprint{\hfill\parbox[b]{0.3\hsize}{ }}

\title{
{\it Supplemental material for}\\
``On-Chip Chiral Single-Photon Interface: Isolation and Unidirectional Emission''
}

\author{Lei Tang,$^1${
\begin{tabular}{ll}
\end{tabular}}
Jiangshan Tang,$^1$
Weidong Zhang,$^2$
Guowei Lu,$^2$
Han Zhang,$^3$
Yong Zhang,$^1$$^*$
Keyu Xia,$^1$$^*$
Min Xiao,$^{1,4}$}
\affiliation
{\it
$^1$National Laboratory of Solid State Microstructures, College of Engineering and Applied Sciences, and School of Physics, Nanjing University, Nanjing 210093, China\\
$^2$State Key Laboratory for Mesoscopic Physics $\&$ Collaborative Innovation Center of Quantum Matter, Department of Physics, Peking University, Beijing 100871, China\\
$^3$Collaborative Innovation Center of Advanced Microstructures, and School of Physics, Nanjing University, Nanjing 210093, China\\
$^4$Department of Physics, University of Arkansas, Fayetteville, Arkansas 72701, USA\\
}

\date{\today}

\maketitle

This supplementary material will provide more details for numerical simulations of the resonator, the chiral light-emitter interaction, the single-photon transport, and more explanation for the the time-energy entanglement. \newline

 \section{Design of a chiral microring resonator}
 In this section, we design a microring resonator on a silica (Si$\text{O}_2$) wafer with an optical chirality (OC) and evaluate its performance through numerical simulation using the commercial finite-difference time-domain (FDTD) mode solver software (Lumerical MODE solutions, www.lumerical.com). In the simulations, the Perfect Match Layer (PML) boundary conditions were applied in both $x$ and $y$ axes; metal boundary conditions were applied in $z$ axes. A mesh size of $10~\nano\meter$ for optical structure region was utilized. We consider a TE mode input. The simulation time step is chosen to be $0.023~\femto\second$.

\begin{figure}[h]
  \centering
  \includegraphics[width = 0.98\linewidth]{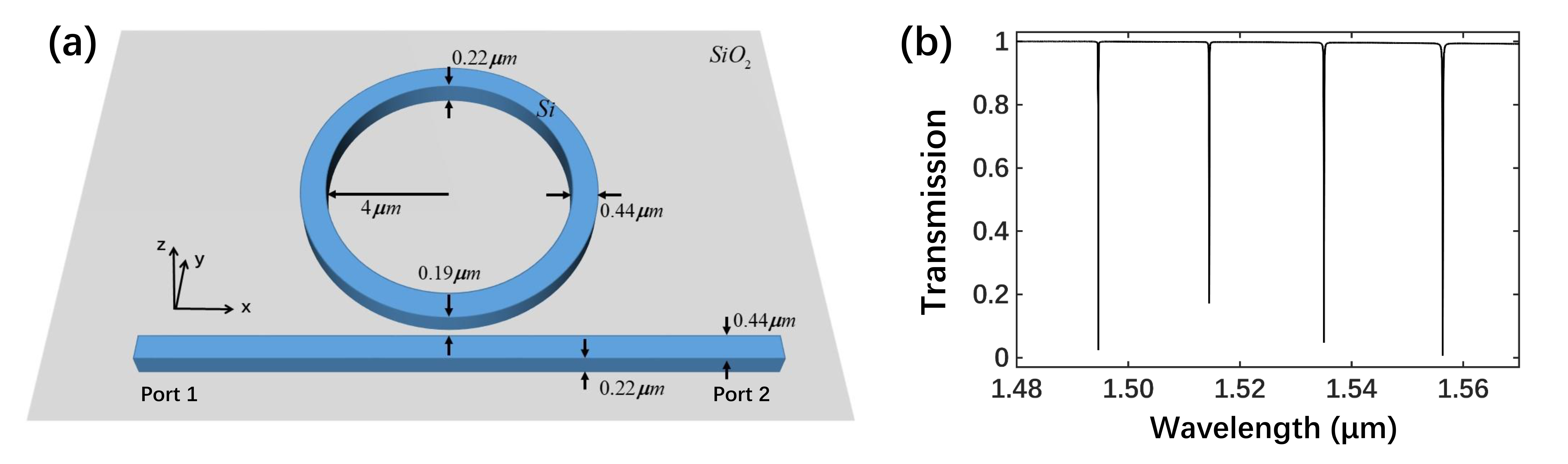}
  \caption{(color online). (a) Structure for the waveguide-resonator device on a silica wafer. (b) Transmission spectrum of the microring resonator.}
  \label{fig:FigS1}
  \end{figure}

\begin{figure}[h]
  \centering
  \includegraphics[width = 0.8\linewidth]{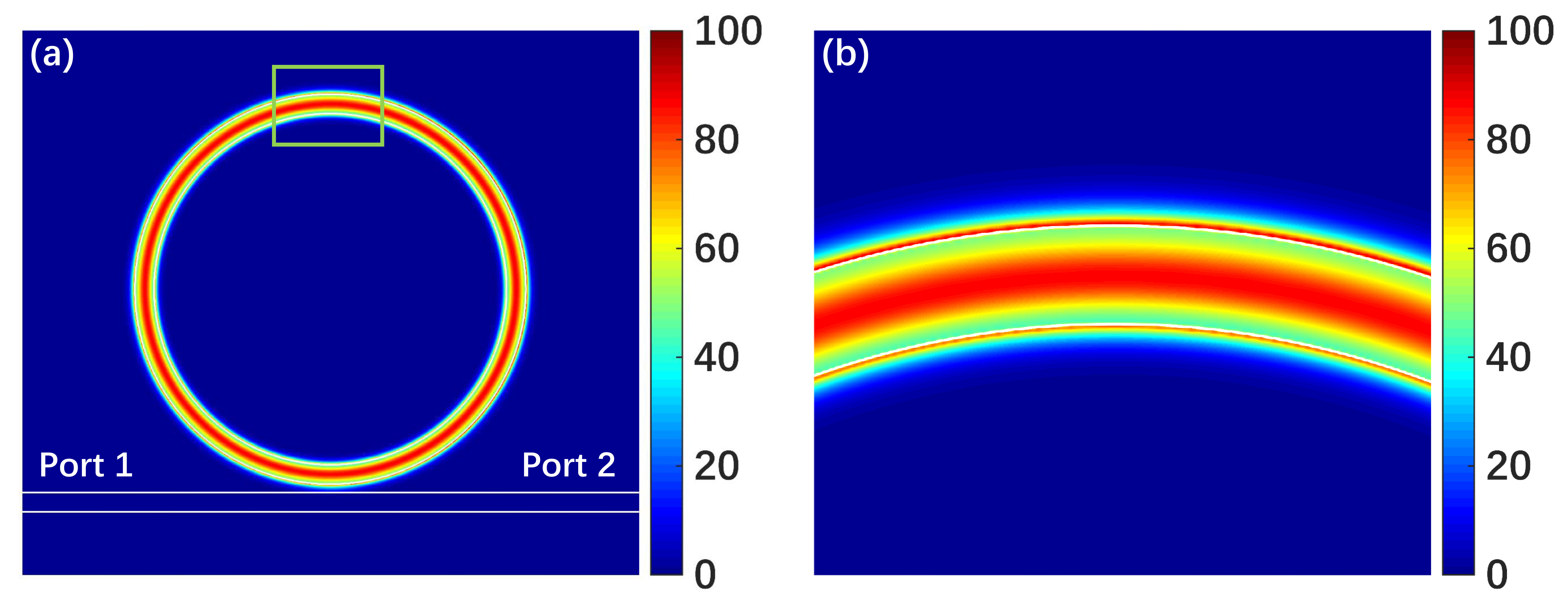}
  \caption{(color online). Intensity distribution ${{{\left| {{\bf{E}}({\bf{r}})} \right|}^2}}$ of the resonator mode at 1.556 $\mu m$ wavelength in the microring resonator. White lines and circles indicate the boundaries of the waveguide and the resonator. (a) The intensity distribution in the whole resonator system. (b) A zoom-in profile within the green box (a).}
  \label{fig:FigS2}
  \end{figure}

 We focus on the OC and the strength of the evanescent electric (e-) field close to the side surface perpendicular to the plane. The evanescent e-field becomes stronger and the OC increases as the cross section of the resonator decreases. To obtain a large OC and a strong evanescent e-field, the cross section of the resonator is designed to have a subwavelength width. We are interested in a resonance mode group around the communication wavelength, i.e. $\lambda \sim 1550~\nano\meter$. In this, the material silicon is transparent for all involved resonator modes. Through numerical simulation, we optimize the waveguide-resonator device for the width. The structure of the device is shown in FIG.~\ref{fig:FigS1}(a). The resonator and the waveguide have the same width $w=440~\nano\meter$ and height $h=220~\nano\meter$. The radius of the resonator is $R$. The gap between the resonator and the waveguide is $G=190~\nano\meter$. We first find a resonance mode at $\lambda \approx 1556~\nano\meter$ for $R= 4.22~ \micro\meter$. This resonance mode can be seen from the dip of the transmission spectrum at $1556~\nano\meter$, shown in FIG.~\ref{fig:FigS2}(a). The vanishing small transmission at this mode indicates that the external decay rate $\kappa_\text{ex}$ due to the loss to the waveguide is equal to the intrinsic decay rate $\kappa_\text{i}$ of the resonator. This means that the critical coupling regime is achieved when the waveguide separates from the resonator by $G=190~\nano\meter$. The total decay rate is $\kappa = \kappa_\text{ex} + \kappa_\text{i}$.

 Part of the transmission spectrum of the microring resonator is shown in FIG.~\ref{fig:FigS1}(b). Here we show the transmissions of four modes. The  mode with an on-resonance wavelength $\lambda_\text{c} \approx 1.556~\micro\meter$, corresponding to the resonance frequency $\omega_\text{c} \approx 2\pi \times 192.67~\tera\hertz$, is under the critical coupling condition and thus its transmission spectrum is nearly zero. The overall quality (Q) factor of the resonator mode can be calculated from the half-height full width, i.e. the linewidth, of the transmission. It is determined by the total decay rate. Under the critical coupling condition, the intrinsic  quality factor $Q_\text{in}$ is twice this overall Q factor. Thus, we find $Q_\text{in}\sim 3.9\times 10^4$ by calculating the linewidth of the transmission spectrum. The corresponding intrinsic decay rate of this resonator mode is thus $\kappa_\text{i} = \omega_\text{c}/Q_\text{in} \approx 2\pi \times 4.94~\giga\hertz$. Through the investigation below, we focus on this mode at $\lambda_\text{c} = 1.556~\micro\meter$.

 The profile of this mode is shown in FIG.~\ref{fig:FigS2}. It is clear that the evanescent e-field along both the outside and inside surface is greatly enhanced in comparison with the conventional bottle-shaped chiral resonator. Hence, it is comparable with the maximal value of the e-field in the middle of the resonator [see FIG.~\ref{fig:FigS2}(b)]. It can be seen from the smooth spatial distribution of the e-field in FIG.~\ref{fig:FigS2} and the singlet dip of the transmission spectrum at $\sim 1.556~\micro\meter$ in FIG.~\ref{fig:FigS1}(b) that the intermode backscattering is very small and can be neglected.

\begin{figure}[h]
  \centering
  \includegraphics[width = 0.8 \linewidth]{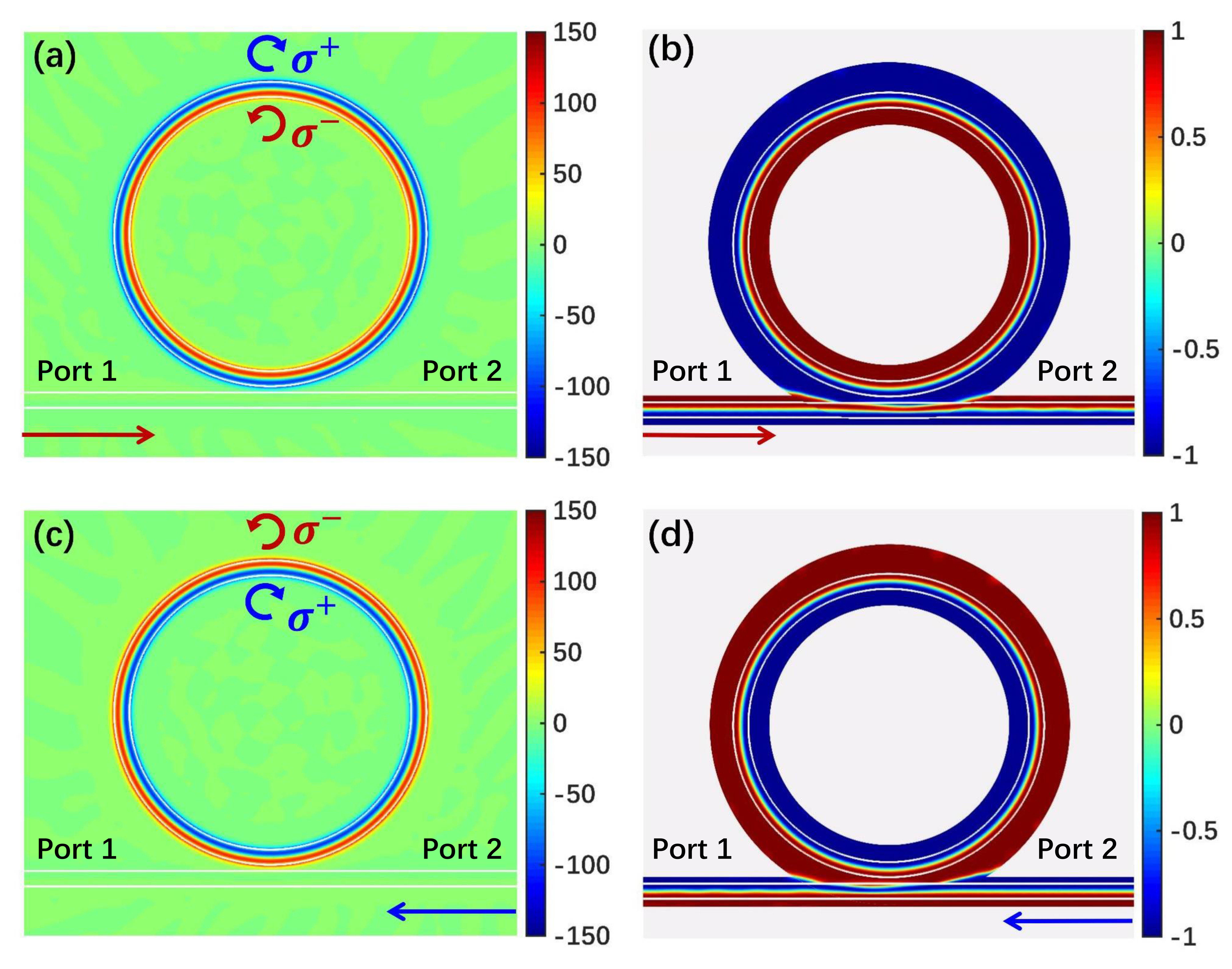}
  \caption{
  (color online). Chiral e-field distributions. (a) and (b) Intensity difference $C$ between the $\sigma^-$- and $\sigma^+$-polarized components of the e-field and the OC $D$ for light input to the port 1. (c) and (d) as for (a) and (b) but for light input to the port 2.}
  \label{fig:FigS3}
  \end{figure}

  Now we go to calculate the OC of the evanescent e-field.
  The OC of the resonator mode is illustrated in FIG.~\ref{fig:FigS3} for light input to the ports 1 and 2. To calculate the OC, we first divide the e-field into two orthogonal components, the $\sigma^-$- and $\sigma^+$-polarized components as ${{\bf{E}}^-}({\bf{r}}) = \bf{E}(\bf{r}) \cdot {{\bf{e}}_{{\sigma ^ - }}}$ and ${{\bf{E}}^+}({\bf{r}}) = \bf{E}(\bf{r}) \cdot {{\bf{e}}_{{\sigma ^ + }}}$ with ${{\bf{e}}_{{\sigma ^ - }}}=({{\bf{e}}_x} - i{{\bf{e}}_y})/\sqrt 2$ and ${{\bf{e}}_{{\sigma ^ + }}}=({{\bf{e}}_x} + i{{\bf{e}}_y})/\sqrt 2$, where ${{\bf{e}}_x}$ and ${{\bf{e}}_y}$ are unit vectors along the $x$ and $y$ directions, respectively. We first plot the intensity difference, $C = (|{{\bf{E}}^-}({\bf{r}})|^2 - |{{\bf{E}}^+}({\bf{r}})|^2)$, in the left panel of FIG.~\ref{fig:FigS3}. It can be seen, for our special design, that the $\sigma_+$ polarization is dominant along the outside surface of the resonator when the light incidents to the system from the port 1 and excites the counter-clockwise (CCW) mode. For light input to the port 2 and exciting the clockwise (CW) mode, the polarization of the evanescent e-field is reversed because of the relation $\bf{E}_{-\bf{k}} (\bf{r}) = \bf{E}^*_{\bf{k}} (\bf{r})$. Obviously, the spin of the resonator mode is locked to the propagation direction of the light. Thus, our design resonator is chiral and possesses the so-called spin-momentum locking. After that, we calculate the OC, defined as $D = (|{{\bf{E}}^-}({\bf{r}})|^2 - |{{\bf{E}}^+}({\bf{r}})|^2)/(|{{\bf{E}}^-}({\bf{r}})|^2 + |{{\bf{E}}^+}({\bf{r}})|^2)$ \cite{PRL.104.163901, PRA.83.021803, PRL.121.043901}, of the resonator mode for two cases of light input to the ports 1 and 2. Obviously, we have $-1\leq D \leq 1$. As shown in FIG.~\ref{fig:FigS3}, the OC is nearly $-1$ for a left input, while it becomes $1$ for an opposite input to the port 2. Here, we only show the OC around the resonator and the waveguide where we are interested in. Therefore, our resonator has a nearly perfect evanescent e-field.  This is the basis of our proposal for the chiral single-photon interface.

\section{Light-emitter interaction}
 We consider a quantum system where a negatively charged quantum dot (QD) with spin-selective transitions is nearby the resonator and couples to the resonator mode. In this section, we estimate the available light-QD interaction strength, which is determined by the dipole moment ${\bf d}$ of the QD and the strength of the evanescent e-field.

 The effective volume of the fundamental mode of the resonator is evaluated by \cite{Phys.Rev.A.75.69}
  \begin{equation}
  \tag{S1}
  \label{eq:eqSl}
   V_\text{m} = \frac{{\int {dV\varepsilon \left( {\bf{r}} \right){{\left| {{\bf{E}}({\bf{r}})} \right|}^2}} }}{{\max \left( {\varepsilon \left( {\bf{r}} \right){{\left| {{\bf{E}}({\bf{r}})} \right|}^2}} \right)}} \;,
  \end{equation}
  where $\epsilon(\bf{r})$ is the electric permittivity of the material at position $\bf{r}$.  According to our numerical simulation, the mode volume of our resonator at $\lambda_\text{c} = 1.556~\micro\meter$  is about $1.55~\micro\meter^{3}$. Correspondingly, the strength of the zero-point fluctuation of this mode is $E_0 = \sqrt{\frac{\hbar\omega_\text{c}}{2\epsilon_0 V_\text{m}}} \approx 6.82\times10^{4}~V/m$, where $\epsilon_0$ is the vacuum permittivity and $\hbar$ is the Planck constant. In start contrast to the conventional Whispering-Gallery mode resonator with a large cross section, the evanescent e-field in our device is close to the maximal amplitude. Thus, it is reasonable to assume that the strength of the e-field coupling to the QD is $E_0$.

 \begin{figure}[h]
  \centering
  \includegraphics[width = 0.98 \linewidth]{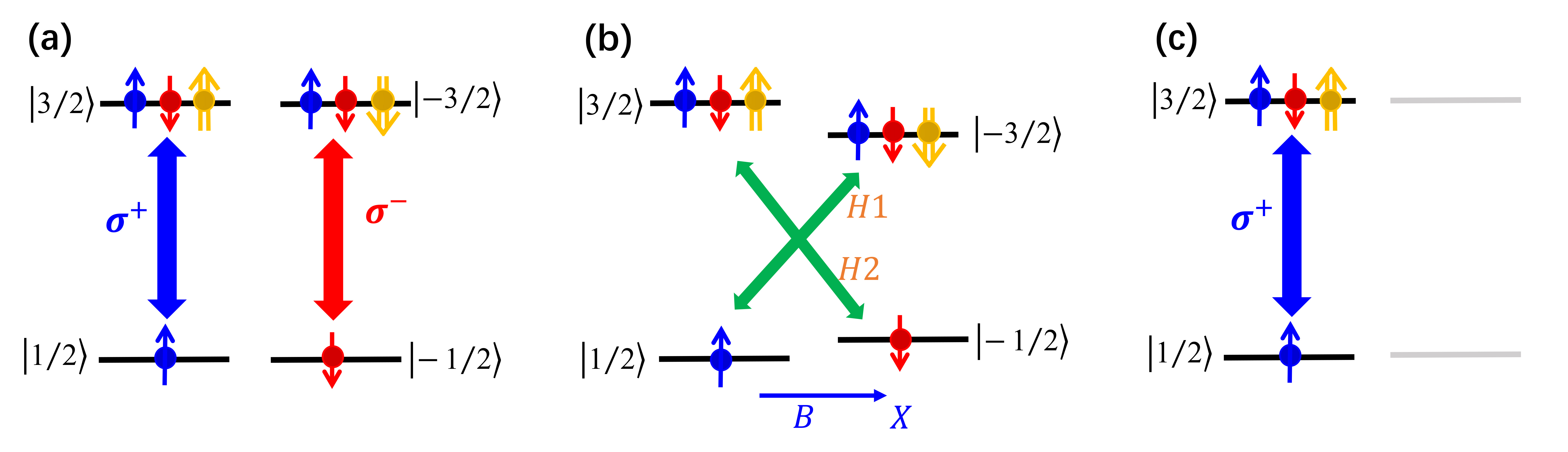}
  \caption{
  (color online). Energy level diagrams for a negatively charged QD with a single electron. (a) Four-level configuration in the absence of a magnetic field.  (b) Four-level configuration with dipole-allowed transitions, enabled by a magnetic field along the $X$ direction. (c) The trion system which has been pumped with linearly polarized light at the magnetic field can be considered as a two-level structure only with ${\sigma ^ + }$ polarized light excitation at zero magnetic field.
  }
  \label{fig:FigS4}
  \end{figure}

  In FIG.~\ref{fig:FigS4}(a), in the absence of a magnetic field, the dipole-allowed transition can only occur between  the spin ground state $\left| {1/2} \right\rangle (\left| { - 1/2} \right\rangle )$ and the trion state $\left| {3/2} \right\rangle (\left| { - 3/2} \right\rangle )$, driven by a ${\sigma ^ + }({\sigma ^ - })$-polarized light. The spin-flip Raman transitions, $|1/2\rangle \leftrightarrow |-3/2 \rangle$ ($| -1/2\rangle \leftrightarrow |3/2\rangle$), are dipole-forbidden. Thus, this QD can be treated as a doubled two-level system without an external magnetic field.

  In FIG.~\ref{fig:FigS4}(b), by applying a magnetic field along the direction perpendicular to the grown direction, the spin-flip Raman transitions are enabled, and can couple to a linearly-polarized e-field, i.e. a $V$ or $H$ polarized field.  In this case, the spin ground state $|1/2\rangle$ or $|-1/2\rangle$ can be selectively prepared with a nearly-unit possibility \cite{QDpreparation1,QDpreparation2,QDpreparation3}. In FIG.~\ref{fig:FigS4}(c), once the spin ground state, e.g. $|1/2\rangle$, is populated, the QD can be treated as a two-level system with a dipole moment interacting with a $\sigma^+$-polarized e-field. After initialization, we can switch off the magnetic field.

\begin{figure}[h]
  \centering
  \includegraphics[width = 0.5 \linewidth]{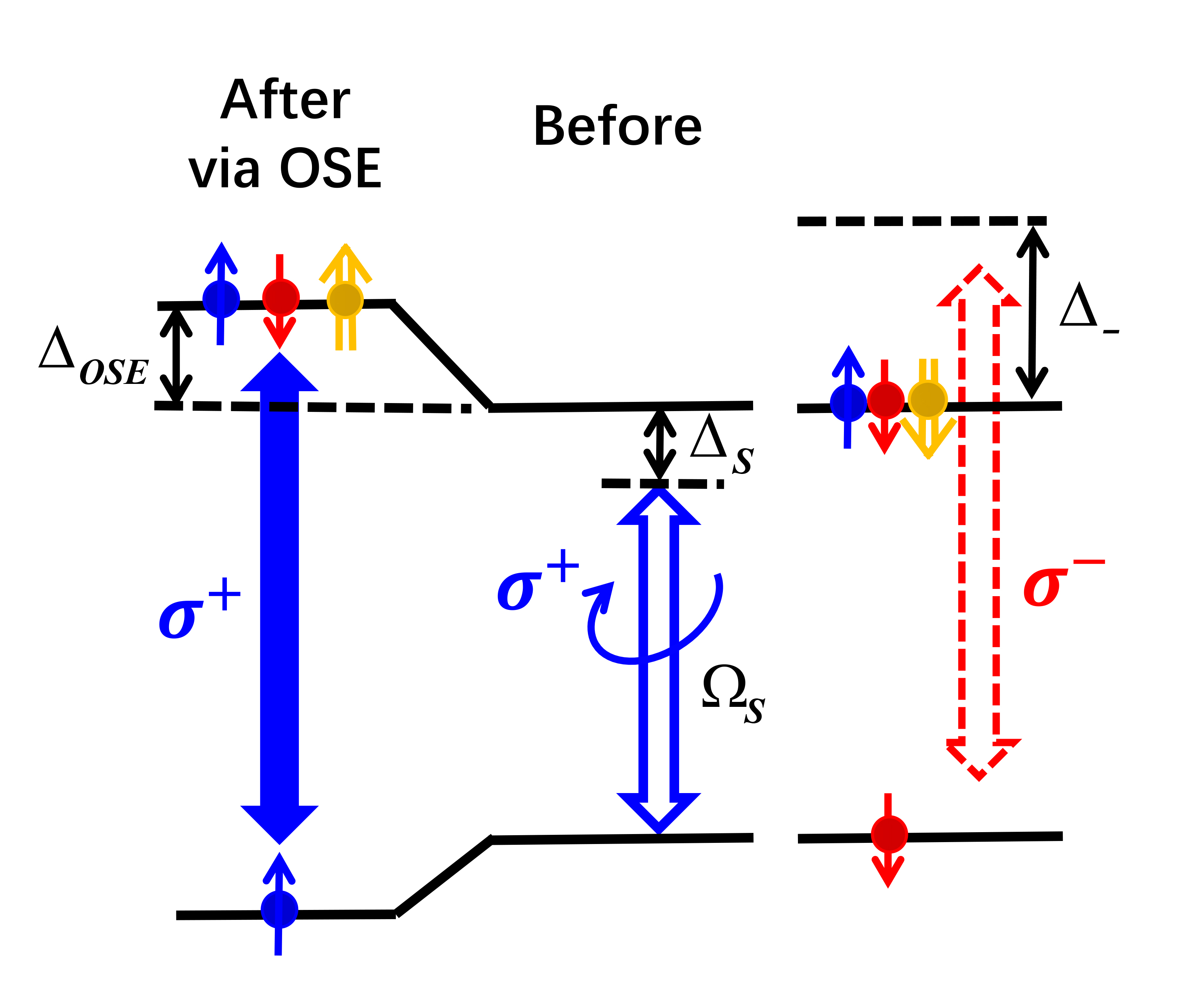}
  \caption{
  (color online). Creating a polarization-selective transition in the QD via the OSE. A $\sigma^+$-polarized classical laser $\Omega_\text{s}$ with a detuning $\Delta_\text{s}$ from the $\sigma^+$-polarized transition $|1/2\rangle \leftrightarrow |3/2\rangle$ is applied to shift the transition energy by $\Delta_\text{OSE}=2\frac{\Omega_\text{s}^2}{\Delta_\text{s}}$. The $\sigma^-$-polarized WGM decouples from the QD because it is detuned by $\Delta_-$ from the relevant transition $|-1/2\rangle \leftrightarrow |-3/2\rangle$.
  }
  \label{fig:FigS5}
  \end{figure}

  Alternatively, we can break the energy degeneracy of transitions $|1/2\rangle \leftrightarrow |3/2\rangle$ and $|-1/2\rangle \leftrightarrow |-3/2\rangle$ by inducing a large optical Start shift with a large detuned circularly-polarized laser \cite{Nat.Commun.6.7665, Nat.Phys.14.1092, Nat.Phys.13.26, Sci.Adv.2.e1600477, Stark2, Stark3}, as depicted in FIG.~\ref{fig:FigS5}. We consider that the $\sigma^+$-polarized transition is shifted by a $\sigma^+$-polarized classical laser to be on-resonance with the CCW WGM. The $\sigma^+$-polarized transition of the QD decouples to the resonator due to a large detuning $\Delta_- = \Delta_\text{c} +2\Delta_\text{OSE}$. In doing so, we can also treat the QD as a two-level system with only $\sigma^+$-polarization-driven transition. Importantly, this protocol allows an all-optical single-photon isolation.
  %

  Self-assembled quantum dots can be engineered to possess a transition at $1.556~\micro\meter$ and their dipole moment can vary from a few Debye to $40$ Debye \cite{PRL.88.087401}. Here, we chose the dipole moment ${\bf d}$ of the QD to be $|{\bf d}|=20$~Debye. We assume that the QD is on resonance with the resonator mode at $1.556~\micro\meter$. Thus, the light-QD coupling rate is
  \begin{equation}
  \tag{S2}
  \label{eq:eqS2}
  g = \frac{{\bf d} \cdot {\bf E}_0}{\hbar}  = 2\pi \times 6.86~\giga\hertz \;.
  \end{equation}

 The decay rate $\gamma_\text{q}$ of the QD is determined by its resonance frequency $\omega_\text{q}$ and dipole moment ${\bf d}$ according to the relation \cite{quantumoptics,Luis.nanofiber}
 \begin{equation}
  \tag{S3}
  \label{eq:eqS3}
   {\gamma _\text{q}} = \frac{{{|{\bf d}|^2}{\omega^3_\text{q}}}}{{3\pi {\varepsilon _0}\hbar {c^3}}} \;,
  \end{equation}
  where $c$ is the light speed in vacuum. Setting $|{\bf d}|=20$ Debye and $\omega_\text{q} = \omega_\text{c}$, we obtain $\gamma_\text{q} = 2\pi\times 5.29~\mega\hertz$.

\medskip

  \section{Single-photon transport model in real space}
  We solve the steady-state transmission using the single-photon transport model in real space, developed by Shen and Fan\cite{Fan}.  The Hamiltonian $H$ below describes the motion of system, consisting of a waveguide, a whispering-gallery mode (WGM) resonator and a two-level system (prepared QD). The Hamiltonian $H$ modeling the single-excitation system is given by
  \begin{equation}
  \tag{S4}
  \label{eq:eqS4}
  \begin{split}
   H/\hbar=& \int {dxc_\text{R}^\dag } \left( x \right)\left( {{\omega _0} - i{v_\text{g}}\frac{\partial }{{\partial x}}} \right){c_\text{R}}\left( x \right) + \int {dxc_\text{L}^\dag } \left( x \right)\left( {{\omega _0} + i{v_\text{g}}\frac{\partial }{{\partial x}}} \right){c_\text{L}}\left( x \right) \\
       & + \left( {{\omega _\text{c}} - i{\kappa_\text{i}}} \right){a^\dag }a + \left( {{\omega _\text{c}} - i{\kappa_\text{i}}} \right){b^\dag }b + \left( {{\Omega _\text{e}} - i{\gamma _\text{q}}} \right)a_\text{e}^\dag {a_\text{e}} + {\Omega _\text{g}}a_\text{g}^\dag {a_\text{g}} \\
       & + \int {dx\delta \left( x \right)\left( {{V_\text{a}}c_\text{R}^\dag \left( x \right)a + V_\text{a}^*{a^\dag }{c_\text{R}}\left( x \right)} \right)} \\
       & + \int {dx\delta \left( x \right)\left( {{V_\text{b}}c_\text{L}^\dag \left( x \right)b + V_\text{b}^*{b^\dag }{c_\text{L}}\left( x \right)} \right)} \\
       & + {g_\text{a}}a{S_ + } + g_\text{a}^*{a^\dag }{S_ - } + {g_\text{b}}b{S_ + } + g_\text{b}^*{b^\dag }{S_ - } \\
       & + h{b^\dag }a + {h^*}{a^\dag}b \;,
  \end{split}
  \end{equation}
  where ${c_\text{R/L}^\dag \left( x \right)}$ is a Bosonic operator creating a right- or left-moving photon at $x$; ${{a^\dag }}$ and ${{b^\dag }}$ are the creation operators for the counterclockwise(CCW) and clockwise(CW) WGM, respectively; Both have the same frequency ${{\omega _\text{c}}}$, $a_\text{e}^\dag (a_\text{g}^\dag)$ is the creation operator of the excited (ground) state of the two-level system; ${S_ + } = a_\text{e}^\dag {a_\text{g}}$ (${S_ - } = a_\text{g}^\dag {a_\text{e}}$) is the raising (lowering)  operator; ${{V_\text{a/b}}}$ is the waveguide-resonator coupling strength of each WGM; ${{g_\text{a/b}}}$ is the QD-resonator coupling strength for each respective WGM; ${\kappa_\text{i}}$ and ${\gamma _\text{q}}$ are the intrinsic decay rate of the resonator and the relaxation rate of the QD, respectively; ${{v_\text{g}}}$ is the group velocity of the photon in the waveguide; ${\Omega _\text{e}} - {\Omega _\text{g}}$ ($\equiv \omega_\text{q}$) is the QD transition frequency; $h$ is the intermode backscattering strength.

  A general state for the system takes the form as
  \begin{equation}
  \tag{S5}
  \label{eq:eqS5}
  \begin{split}
     \left| \psi  \right\rangle  = \int {dx\left[ {{{\widetilde \phi }_\text{R}}\left( {x,t} \right)c_\text{R}^\dag \left( x \right) + {{\widetilde \phi }_\text{L}}\left( {x,t} \right)c_\text{L}^\dag \left( x \right)} \right]} \left| \varnothing  \right\rangle  + \left[ {{{\widetilde e}_\text{a}}\left( t \right){a^\dag } + {{\widetilde e}_\text{b}}\left( t \right){b^\dag } + {{\widetilde e}_\text{q}}\left( t \right){S_ + }} \right]\left| \varnothing  \right\rangle,
  \end{split}
  \end{equation}
  associated with the eigenvalue $\omega$, so that $\widetilde X = {e^{ - i\omega t}}X$ with $X \in \left\{ {{\phi _\text{R}},{\phi _\text{L}},{e_\text{a}},{e_\text{b}},{e_\text{q}}} \right\}$. ${{{\widetilde \phi }_\text{R/L}}\left( {x,t} \right)}$ is the single-photon wave function of the right- or left-moving mode; ${{{\widetilde e}_\text{a/b}}}$ is the excitation amplitude of each respective WGM; ${{{\widetilde e}_\text{q}}}$ is the excitation amplitude of the QD. $\left| \varnothing  \right\rangle$ is the vacuum state, which has zero photon in the system and with the QD in the ground state. For the purpose of solving the transmission amplitude of an incident single photon, we take ${\phi _\text{R}}\left( x \right) = {e^{iqx}}\left[ {\theta \left( { - x} \right) + {t_ + }\theta \left( x \right)} \right]$, ${\phi _\text{L}}\left( x \right) = {r_ + }{e^{ - iqx}}\theta \left( { - x} \right)$ for a left-hand (right-moving) incident photon and ${\phi _\text{L}}\left( x \right) = {e^{ - iqx}}\left[ {\theta \left( x \right) + {t_ - }\theta \left( { - x} \right)} \right]$, ${\phi _\text{R}}\left( x \right) = {r_ - }{e^{iqx}}\theta \left( x \right)$ for a right-hand (left-moving) input at location $x$\cite{Opt.Lett.30.2001,Xia.2014}, where $q$ is the wave vector of the input field around the frequency $\omega$, ${{t_ \pm }}$ is the transmission amplitude, and ${{r_ \pm }}$ is the reflection amplitude. $\theta \left( x \right)$ is the Heaviside step function that $\begin{array}{l}\theta \left( x \right)\left| {_{x = 0} = 1/2} \right.\\\end{array}$, $\frac{{\partial \theta \left( x \right)}}{{\partial x}}\left| {_{x \to {0_ + }} = 1} \right.$, $\frac{{\partial \theta \left( x \right)}}{{\partial x}}\left| {_{x \to {0_ - }} =  - 1} \right.$ \cite{Fan,Xia.2014}.

  Based on the Schr\"odinger equation $i\hbar \frac{\partial }{{\partial t}}\left| \psi  \right\rangle  = H\left| \psi  \right\rangle$ in the real space, we can derive the steady-state solution that for transmission amplitudes. In our case,  the photon only drives the ${\sigma ^ + }$ transition of the QD, and we have $\left| {{V_\text{a}}} \right| = \left| {{V_\text{a}}} \right| \equiv V$.  The decay rate of the resonator due to the external coupling $V$ to the waveguide is given by $\kappa_\text{ex}  \equiv {{{V^2}} \mathord{\left/{\vphantom {{{V^2}} {2{v_\text{g}}}}} \right.\kern-\nulldelimiterspace} {2{v_\text{g}}}}$.

  To consider a general case in which the evanescent field of the ring resonator is not perfectly circularly polarized, we define $\alpha  = \left| {{{\bf{E}}_\text{CCW}} \cdot {{\bf{e}}_{{\sigma ^ + }}}} \right|/\left| {{{\bf{E}}_\text{CCW}}} \right|$ and $\beta  = \left| {{{\bf{E}}_\text{CCW}} \cdot {{\bf{e}}_{{\sigma^- }}}} \right|/\left| {{{\bf{E}}_\text{CCW}}} \right|$ with ${\alpha ^2} + {\beta ^2} = 1$. $\alpha$ and $\beta$ indicate the projection of the field to the unit vector ${{{\bf{e}}_{{\sigma^+ }}}}$ and ${{{\bf{e}}_{{\sigma^- }}}}$, respectively. The modes of the ring resonator are time-reversal symmetric. Therefore, we have ${\omega _\text{CCW}}={\omega _\text{CW}}$ and ${{{\bf{E}}_\text{CCW}}({\bf{r}})={\bf{E}}_\text{CW}^*({\bf{r}})}$, implying that $\left| {{{\bf{E}}_\text{CW}} \cdot {{\bf{e}}_{{\sigma ^ - }}}} \right|/\left| {{{\bf{E}}_\text{CW}}} \right| = \alpha$ and $\left| {{{\bf{E}}_\text{CW}} \cdot {{\bf{e}}_{{\sigma ^ + }}}} \right|/\left| {{{\bf{E}}_\text{CW}}} \right| = \beta$, and yielding $D = (\left| {{{\bf{E}}_\text{CCW}} \cdot {{\bf{e}}_{{\sigma^- }}}} \right|^2 - \left| {{{\bf{E}}_\text{CCW}} \cdot {{\bf{e}}_{{\sigma^+ }}}} \right|^2)/(\left| {{{\bf{E}}_\text{CCW}} \cdot {{\bf{e}}_{{\sigma^- }}}} \right|^2 + \left| {{{\bf{E}}_\text{CCW}} \cdot {{\bf{e}}_{{\sigma^+ }}}} \right|^2)=(\beta^2-\alpha^2)/(\beta^2+\alpha^2)$.
 Thus, we can obtain $\alpha=\sqrt{(1-D)/2}$ and $\beta=\sqrt{(1+D)/2}$.
 The evanescent field ${\bf{E}}_\text{CCW}$(${\bf{E}}_\text{CW}$) of the CCW(CW) mode drives the ${{\sigma ^ + }}$-polarized transition of the QD. The coupling strengths are $|{g_\text{a}}| \propto \left| {{{\bf{d}}} \cdot ({\bf{E}}_\text{CCW}}\cdot {{\bf{e}}_{{\sigma^+ }}}) \right|$ and $|{g_\text{b}}| \propto \left| {{{\bf{d}}} \cdot ({\bf{E}}_\text{CW}}\cdot {{\bf{e}}_{{\sigma^+ }}}) \right|$, respectively, where ${{{\bf{d}}}}$ is the electric dipole moment of the QD. In this, we get $\left| {{g_\text{a}}} \right| = \alpha g$, $\left| {{g_\text{b}}} \right| = \beta g$. For instance, $D = -1$ corresponding to $\alpha =1$ and $\beta =0$; $D= -0.99$ corresponding to $\alpha =\sqrt {0.995} $ and $\beta =\sqrt{0.005}$.

\begin{figure}[h]
  \centering
  \includegraphics[width = 0.98 \linewidth]{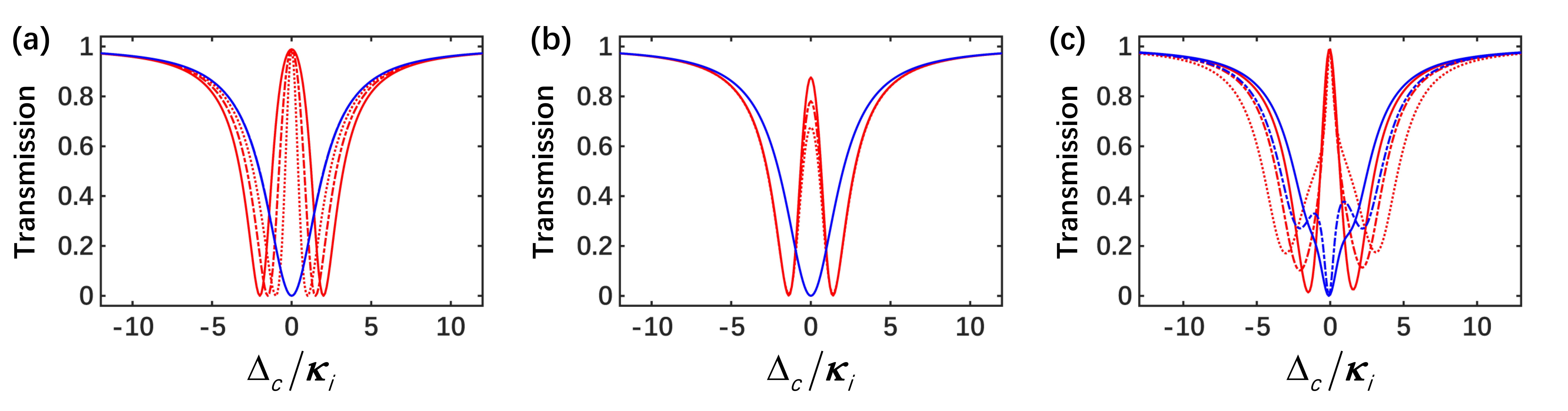}
  \caption{
  (color online). Steady-state transmissions. Red (blue) curves are for the forward-moving (backward-moving) transmissions ${T_ +}$ ($T_-$). (a) The steady-state transmission for $|D| = 0.99$, $\gamma_\text{q} = 2\pi  \times 5.29~\mega\hertz$ and $|h|=0$. Solid, dashed and dotted curves are for $g=2\kappa_\text{i}$, $1.5\kappa_\text{i}$ and $\kappa_\text{i}$, respectively.
  (b) The steady state transmission for $|D| = 0.99$, $g=2\pi\times 6.86 ~\giga\hertz$ and $|h|=0$. Solid, dashed and dotted curves are for $\gamma_\text{q}/2\pi=300~\mega\hertz$, $600~\mega\hertz$ and $1~\giga\hertz$, respectively.
  (c) The steady state transmission for $|D| = 0.99$, $g=2\pi\times 6.86 ~\giga\hertz$ and ${\gamma_\text{q}} = 2\pi  \times 5.29~\mega\hertz$; solid, dashed and dotted curves are for $|h|= \kappa_\text{i}$, $2\kappa_\text{i}$ and $3\kappa_\text{i}$, respectively.
  }
  \label{fig:FigS6}
  \end{figure}

 Using the single-photon transport method, we can find the steady-state transmission, as shown in FIG.~\ref{fig:FigS6}. The bandwidth for photon isolation is crucially dependent on the resonator-QD interaction, but the isolation contrast remains unchanged at resonance, i.e. at $\Delta_\text{c}=0$. It can be seen from FIG.~\ref{fig:FigS6}(a) that the nonreciprocal spectral window becomes narrower and narrower as the resonator-QD coupling strength decreases. For example, the bandwidth for $g=2\kappa_\text{i}, 1.5\kappa_\text{i}, \kappa_\text{i}$ is about $2\pi\times 12.10$~\giga\hertz, $2\pi \times 7.90$~\giga\hertz, and $2\pi \times 3.95$~\giga\hertz.
 The change of the QD decay causes the increase of the insert loss but has little influence on the nonreciprocal bandwidth, see FIG.~\ref{fig:FigS6}(b). The insert loss $\mathscr{L}$ increases from $0.57$~dB  to $1.70$~dB as $\gamma_\text{q}$ increases from $2\pi \times 300$~\mega\hertz~  to $2\pi \times 1$~\giga\hertz. In contrast, the backscattering has a complicate impact on the transmission spectra, see FIG.~\ref{fig:FigS6}(c). Basically, it reduces the bandwidth of photon isolation.

\medskip

\medskip

  \section{Time evolution of single-photon pulses}
  The steady-state solution has shown the isolation when a single photon is injected into either the forward or backward direction.  Here, we check how well our single-photon isolator works when single-photon wavepackets enters the system from two opposite directions simultaneously.  To do so, we numerically simulate the time evolution of a single-photon pulse in the wave vector ($\bf k$) space. We concentrate on the propagation of the single-photon wave packets through the QD-resonator system in the ideal case with a unity OC, i.e. $|D|=1$. We start our discussion from the Hamiltonian of the system in the $\bf k$ space, $H=H_0+H_\text{I}$ \cite{Fan1, Opt.Lett.30.2001}:
  \begin{subequations}
  \begin{align}
  \label{eq:eqS6}
  \tag{S6a}
  &H_0/\hbar= \int{d{\bf k}\omega_{\bf k} c^\dag_{\bf k} c_{\bf k}}+\int{d{\bf k}\omega_{\bf k} d^\dag_{\bf k} d_{\bf k}}+(\omega_\text{c}-i\kappa_\text{i})a^\dag a + (\omega_\text{c}-i\kappa_\text{i})b^\dag b + (\omega_\text{q}-i\gamma_\text{q})a^\dag_\text{e} a_\text{e}\;,\\
  \tag{S6b}
  &H_\text{I}/\hbar=\int{d{\bf k}V_{\bf k}(c^\dag_{\bf k} a+a^\dag c_{\bf k})}+\int{d{\bf k}V_{\bf k}(d^\dag_{\bf k} b+b^\dag d_{\bf k})}+g(S_+ a+a^\dag S_-)\;,
  \end{align}
  \end{subequations}
  where $H_0$ is the Hamiltonian of the system, including the waveguide, ring resonator and the QD, $H_\text{I}$ is the Hamiltonian of the interaction between the waveguide and the ring resonator, and the QD-resonator interaction; $c_{\bf k}^\dag$  $(d_{\bf k}^\dag)$ is the creation operator for the right- or left-moving photon with a wave vector $\bf k$; $V_{\bf k}$ is the coupling strength the resonator mode and the waveguide mode with a wave vector $\bf k$; the definitions of $a^\dag$, $a_e^\dag$, $S_+$, $S_-$, $\omega_c$, $\omega_q$, $\kappa_\text{i}$, $\gamma_\text{q}$ and $g$ are the same as in section III.

  For an arbitrary frequency $\omega_0$ of a propagating waveguide mode that is away from the cutoff of the dispersion, with the corresponding wave vector $\pm {\bf k}_0$, one can approximate $\omega_{\bf k}$ around ${\bf k}_0$ and $-{\bf k}_0$ as \cite{Fan1}
  \begin{subequations}
  \begin{align}
  \label{eq:eqS7}
  \tag{S7a}
  &\int_{{\bf k}\simeq {\bf k}_0}dk \omega_{\bf k}c^\dag _{\bf k} c_{\bf k}\simeq \int_{{\bf k}\simeq {\bf k}_0}d{\bf k}[\omega_0+v_\text{g}({\bf k}-{\bf k}_0)]c^\dag _{\bf k} c_{\bf k}\;,\\
  \tag{S7b}
  &\int_{{\bf k}\simeq -{\bf k}_0}dk \omega_{\bf k}c^\dag _{\bf k} c_{\bf k}\simeq \int_{{\bf k}\simeq -{\bf k}_0}d{\bf k}[\omega_0-v_\text{g}({\bf k}+{\bf k}_0)]c^\dag _{\bf k} c_{\bf k}\;.
  \end{align}
  \end{subequations}
 In our system, we define $\Delta_\text{c} =\omega_0-\omega_\text{c}$ and $\Delta_\text{q}=\omega_0-\omega_\text{q}$. When the resonance frequency of the QD is away from the cutoff frequency of the dispersion relation, we can rewrite the Hamiltonian in the $\bf k$ space as
\begin{subequations}
  \begin{align}
  \tag{S8a}
  \label{eq:eqS8a}
  &H_0/\hbar=v_\text{g}\sum_{{\bf k}}({\bf k}-{\bf k}_0)c_{\bf k}^\dag c_{\bf k} - v_\text{g}\sum_{{\bf k}}({\bf k}+{\bf k}_0)d_{\bf k}^\dag d_{\bf k} -(\Delta_\text{c}+i\kappa_\text{i})a^\dag a-(\Delta_\text{c}+i\kappa_\text{i})b^\dag b-(\Delta_\text{q}+i\gamma_\text{q})a_\text{e}^\dag a_\text{e}\;,\\
  \tag{S8b}
  \label{eq:eqS8b}
  &H_\text{I}/\hbar=\sum_{{\bf k}}V_{\bf k}(c_{\bf k}^\dag a+a^\dag c_{\bf k})+\sum_{{\bf k}}V_{\bf k}(d_{\bf k}^\dag b+b^\dag d_{\bf k})+g(S_+a+a^\dag S_-)\;.
  \end{align}
  \end{subequations}
  For a single-photon wave packet, the general state of the system takes the form  \cite{Fan}
 \begin{equation}
  \tag{S9}
  \label{eq:eqS9}
  \begin{split}
     |\psi\rangle=\sum_{{\bf k}}\phi_{\bf k}^\text{c}({\bf k},t)c_{\bf k}^\dag \left| \varnothing  \right\rangle + \sum_{{\bf k}}\phi_{\bf k}^\text{d}({\bf k},t)d_{\bf k}^\dag \left| \varnothing  \right\rangle + e_\text{a}(t)a^\dag\left| \varnothing  \right\rangle + e_\text{b}(t)b^\dag\left| \varnothing  \right\rangle + e_\text{q}(t)S_+\left| \varnothing  \right\rangle \;,
  \end{split}
  \end{equation}
  where $\phi_{\bf k}^\text{c} (\phi_{\bf k}^\text{d})$ is the wave packet of the right- (left-) moving photon appearing at $x$ in the waveguide. $e_\text{a}$ ($e_\text{b}$) is the excitation of the CCW (CW) WGM, and $e_\text{q}$ is the excitation of the QD in the $\sigma^+$-polarized transition, $\left| \varnothing  \right\rangle$ is the vacuum state.

  Substituting Eqs.~S8 and ~\ref{eq:eqS9} into the Schr\"odinger equation $i\hbar \frac{\partial }{{\partial t}}\left| \psi  \right\rangle  = H\left| \psi  \right\rangle$, we get the following set of equations of motion for the propagation of single-photon wave packets in the $\bf k$ space:
 \begin{subequations}
 \begin{align}
  \tag{S10a}
  \label{eq:eqS10a}
  &i{\partial_t}{\phi_{\bf k}^\text{c}({\bf k},t)}=v_\text{g}({\bf k}-{\bf k}_0)\phi_{\bf k}^\text{c}({\bf k},t)+V_{\bf k} e_\text{a}(t)\;,\\
  \tag{S10b}
  \label{eq:eqS10b}
  &i{\partial_t}{\phi_{\bf k}^\text{d}({\bf k},t)}=-v_\text{g}({\bf k}+{\bf k}_0)\phi_{\bf k}^\text{d}({\bf k},t)+V_{\bf k} e_\text{b}(t)\;,\\
  \tag{S10c}
  \label{eq:eqS10c}
  &i{\partial_t}{e_\text{a}(t)}=-(\Delta_\text{c}+i\kappa_\text{i})e_\text{a}(t)+V_{\bf k} \phi_{\bf k}^\text{c}({\bf k},t)+ge_\text{q}(t)\;,\\
  \tag{S10d}
  \label{eq:eqS10d}
  &i{\partial_t}{e_\text{b}(t)}=-(\Delta_\text{c}+i\kappa_\text{i})e_\text{b}(t)+V_{\bf k} \phi_{\bf k}^\text{d}({\bf k},t)\;,\\
  \tag{S10e}
  \label{eq:eqS10e}
  &i{\partial_t}{e_\text{q}(t)}=-(\Delta_\text{q}+i\gamma_\text{q})e_\text{q}(t)+ge_\text{a}(t)\;.
  \end{align}
  \end{subequations}
  Numerical integration of this set of derivative equations can obtain the time-evolution of the photon wave packet. In our case, we need the initial state for the input $\phi_{\bf k}^\text{c}({\bf k},0)$ and $\phi_{\bf k}^\text{d}({\bf k},0)$, which can be found from $\phi_+(x,0)$ and $\phi_-(x,0)$ by applying the Fourier transformation. We assume that $V_{\bf k} = V$ for all $\bf k$ within the band of the input single-photon pulse.
  Here, we assume Gaussian pulse wave packet inputs from both the left-hand and right-hand sides at the same time that $\phi(x,0)=\phi_+(x,0)+\phi_-(x,0)$ where  $\phi_+(x,0)= \pi^{-1/4} \tau_\text{p}^{-1/2}\exp( - {(x - {x_{0_\text{L}}})^2}/2\tau ^2_\text{p})/\sqrt{2}$ and $\phi_-(x,0)= \pi^{-1/4} \tau_\text{p}^{-1/2} \exp( - {(x - {x_{0_\text{R}}})^2}/2\tau ^2_\text{p})/\sqrt{2}$, where $\tau_\text{p}$ is the spatial duration of the pulse, and $x_{0_\text{L}}$ ($x_{0_\text{R}}$) indicates the position away from the resonator in the left-hand (right-hand) side. The input is normalized to include a single excitation, $\int_{ - \infty }^{ + \infty } {{\phi ^*}(x,0)} \phi (x,0)dx = 1$. However, the input is a superposition of single-photon wavepacket with $\int_{ - \infty }^{ + \infty } {{\phi_+^*}(x,0)} \phi_+ (x,0)dx = 1/2$ and $\int_{ - \infty }^{ + \infty } {{\phi_-^*}(x,0)} \phi_- (x,0)dx = 1/2$. We set the velocity of light in the waveguide $v_\text{g}=1$, and choose a number for $V$ that the critical coupling condition holds. Other parameters are $g=1.39 \kappa_\text{i}$ and $\gamma_\text{q}=10^{-3}\kappa_\text{i}$.

  Then we solve the equations Eqs.~\ref{eq:eqS10a} $\sim$ \ref{eq:eqS10e}, and obtain the solution with the state ($\phi_{\bf k}^\text{c}({\bf k},t_m)$, $\phi_{\bf k}^\text{d}({\bf k},t_m)$) as the photon wave packets pass through the QD-resonator system at $t=t_m$. After obtaining these wave packets in the $\bf k$ space, we do the inverse Fourier transformation to get $\phi_+(x,t_m)$ and $\phi_-(x,t_m)$ in real space.
%
  Finally, we obtain time-dependent transport of this single-photon wave packet that is presented in our main texts, see Fig.~3(c).

\begin{figure}[h]
  \centering
  \includegraphics[width = 0.8 \linewidth]{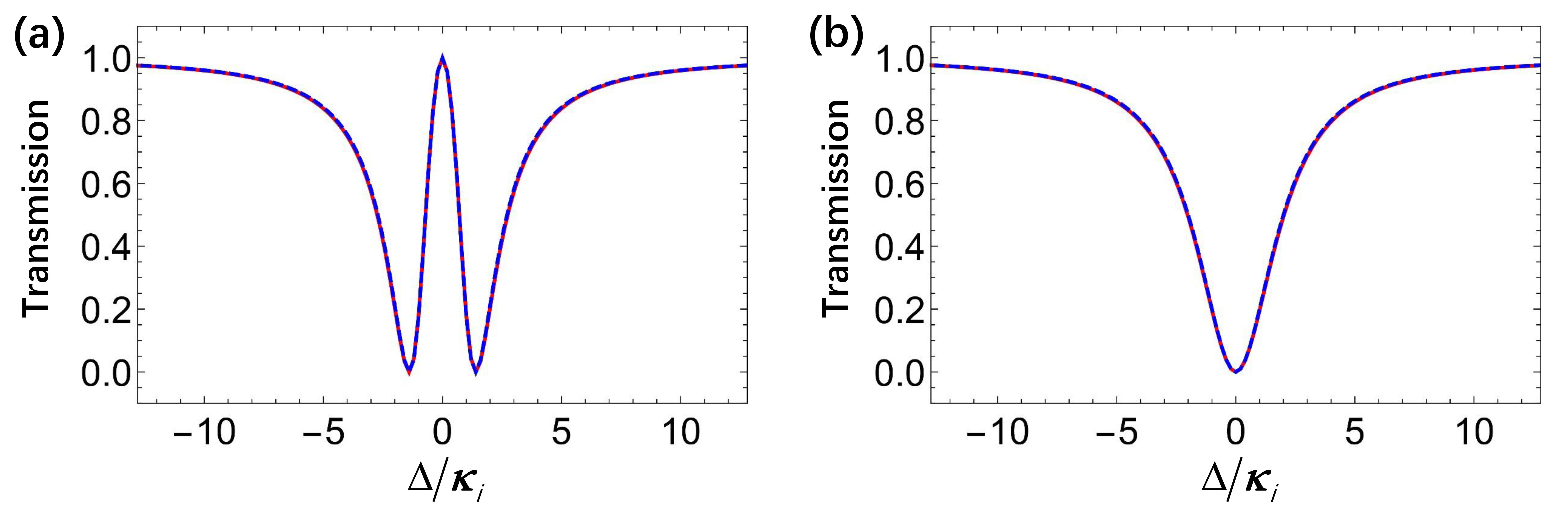}
  \caption{
  (color online). The forward (a) and backward (b) transmissions of the steady-state solution (red dashed curves) and numerical results (blue solid curves).
  }
  \label{fig:FigS7}
  \end{figure}

  We also numerically solve the transmission spectra using the same parameters but a short input single-photon pulse. The numerical results are in excellent agreement with the stead-state solutions, see FIG.~\ref{fig:FigS7}. The numerical and the analytic solutions are completely overlapped.

\section{Influence of a backscattering}

  Here we discuss the influence of the backscattering in the resonator due to some impurity or surface roughness. Basically, the backscattering causes the coupling between the two opposite-propagation modes, i.e. the CW and CCW modes.  For example, the left-hand incident photon excites the CCW mode dominantly. In the presence of a backscattering, the CCW mode couples to the CW mode to some degree. Phenomenally, it results in the splitting of the transmission spectrum. However, a large decay rate of the resonator mode, corresponding to a low Q factor, can significantly suppress the backscattering process. To evaluate the influence of the backscattering on the performance of our device, we consider a backscattering rate $h$, resulting in an interaction Hamiltonian $H_\text{sc} = h a^\dag b + h^* a b^\dag$. We denote the CCW and CW modes $a$ and $b$. These two ring resonator modes typically have the same intrinsic decay rate $\kappa_\text{i}$ and the same external decay rate $\kappa_\text{ex}$ that $\kappa = \kappa_\text{i} + \kappa_\text{ex}$.  They have a degenerate resonance frequency. The quantum Langevin equation describing the motion of these two modes, driven by an external field with a detuning $\Delta_\text{in}$, takes the form
  \begin{subequations}
  \begin{align}
  \tag{S11a}
  \label{eq:eqS11a}
  \dot a & = -i \Delta_\text{in} a - \kappa a + \sqrt {2{\kappa _\text{ex}}} {\alpha _\text{in}} - i h b \;,\\
  \tag{S11b}
  \label{eq:eqS11b}
  \dot b & = -i \Delta_\text{in} a - \kappa b - i h^* a \;,
  \end{align}
  \end{subequations}
  for a where the input $\alpha_\text{in}$ excites the mode $a$.
  We can easily get the steady-state solution as
  \begin{subequations}
  \begin{align}
  \tag{S12a}
  \label{eq:eqS12a}
  a & = \frac{(i \Delta_\text{in} + \kappa) \sqrt{2\kappa_\text{ex} }\alpha_\text{in}}{(i\Delta_\text{in}+\kappa)^2 + |h|^2} \;,\\
  \tag{S12b}
  \label{eq:eqS12b}
  b & =  \frac{-i h^* \sqrt{2\kappa_\text{ex} }\alpha_\text{in}}{(i\Delta_\text{in}+\kappa)^2 + |h|^2} \;.
  \end{align}
  \end{subequations}
  Thus, we have $|b/a|^2 = |h|^2/\kappa^2$ for $\Delta_\text{in}=0$. This means that the influence of the backscattering is very small if $|h| \ll \kappa$. Therefore, the transmission spectrum of a ring resonator with a low Q factor will be robust against the backscattering. In other word, the excitation of its unwanted mode (here it is the CW mode $b$) due to the backscattering is negligible.

\medskip

\section{Time-bin emission}
  We consider an interaction between a
 two-level system (a single prepared QD) and a cavity mode. In our case, the two-level system is a negatively-charged QD prepared in one of the hole-spin excited states. Thus, it emits a single photon into either the CCW mode or the CW mode. The cavity mode has a decay rate of $\kappa=\kappa_\text{i}+\kappa_\text{ex}$ as mentioned above. Before discussing the single-photon emission, we first present a formula for the collection efficient of photons decaying from the cavity with out the QD to the nearby waveguide. Assuming that the bare cavity is prepared in a state with photon number $\langle a^\dag(0)a(0)\rangle$.  The evolution of the bare cavity is governed by the quantum Langevin equation
\begin{equation}
\tag{S13}
\label{eq:QL}
   \dot a(t) = - (\kappa_\text{i} + \kappa_\text{ex}) a(t) \;.
 \end{equation}
The solution is
\begin{equation}
\tag{S14}
   a(t) = e^{-(\kappa_\text{i}+\kappa_\text{ex})t}a(0) \;.
 \end{equation}
 The photon number in the cavity is
 \begin{equation}
 \tag{S15}
   \langle a^\dag(t)a(t)\rangle = e^{-2(\kappa_\text{i}+\kappa_\text{ex})t}\langle a^\dag(0)a(0)\rangle \;.
  \end{equation}
  In the absence of external driving, using the input-output relation $a_\text{out} = \sqrt{2\kappa_\text{ex}}$, the total photon number collected in the waveguide can be calculated as
\begin{equation}
\tag{S16}
  \begin{split}
 n_\text{WG} & =   \int_{\rm{0}}^{{\rm{ + }}\infty } {\left\langle {a_\text{out}^\dag (t){a_\text{out}}(t)} \right\rangle } dt \\
  & = 2{\kappa _\text{ex}}\int_{\rm{0}}^{{\rm{ + }}\infty } {\left\langle {{a^\dag }(t)a(t)} \right\rangle } dt \\
    & = \frac{\kappa_\text{ex}}{\kappa_\text{ex}+\kappa_\text{i}}\;.
  \end{split}
  \end{equation}
  Thus, to collect the photons from the cavity efficiently, we need that the waveguide is overcoupled to the cavity, i.e. $\kappa_\text{ex} \gg \kappa_\text{i}$.

 For the single-photon emission, no external field is input into the system. The Hamiltonian describing the dynamics of the system can be written as
  \begin{equation}
  \tag{S17}
  \label{eq:eqS13}
  \begin{split}
   H = \omega_\text{c} a^{\dag}a + \omega_\text{q} S_+S_- + g(S_+ a + a^\dag S_-) \;,
  \end{split}
  \end{equation}
  where the definitions of $a^\dag$, $S_+$, $S_-$, $\omega_c$, $\omega_q$, and $g$ are the same as in section III.
In our case, we have $\omega_c=\omega_q$ and $\gamma_\text{q} \ll \kappa$. We use the Tan's quantum toolbox \cite{toolbox} to simulate the evolution of the QD-resonator system and calculate the excitation of a single photon entering the waveguide. The numerical result is shown in Figs.~4(e) and (f).

%


%